\documentclass[11pt,a4,twoside]{article}
\usepackage[dvips,a4paper,left=2.5cm,right=2.5cm,top=2.5cm,bottom=2.5cm,headsep=1em]{geometry}
\usepackage{fancyhdr}
\usepackage{titlesec}
\usepackage[latin1]{inputenc}
\usepackage{amsmath,amsfonts,array}
\usepackage{rotating}
\usepackage[font={small}]{caption}
\usepackage{natbib}
\usepackage{lmodern,slantsc}
\usepackage{multirow}
\usepackage{chngcntr}
\usepackage{placeins}
\usepackage{caption}
\usepackage{enumitem}
\usepackage[breaklinks,
colorlinks=true, linkcolor=blue, citecolor=black, urlcolor=blue,
pdfborder={0 0 0}, pdfpagelabels]{hyperref}
\usepackage{graphicx}
\newcommand{\PRLsep}{\noindent\makebox[\linewidth]
{\resizebox{0.3333\linewidth}{1pt}{$\bullet$}}\bigskip}
\usepackage{tikz}
\def\vx{\mathbf x}

\def\vc{\mathbf c}

\def\vf{\mathbf f}
\def\vg{\mathbf g}

\def\vr{\mathbf r}

\def\vu{\mathbf u}
\def\vv{\mathbf v}

\def\vX{\mathbf X}

\def\vZ{\mathbf Z}
\def\vA{\mathbf A}

\def\vC{\mathbf C}
\def\vF{\mathbf F}
\def\vG{\mathbf G}

\def\vJ{\mathbf J}
\def\vL{\mathbf L}

\def\vN{\mathbf N}

\def\vQ{\mathbf Q}
\def\vR{\mathbf R}
\def\vS{\mathbf S}
\def\vT{\mathbf T}

\def\vW{\mathbf W}
\def\vX{\mathbf X}
\def\v0{\boldsymbol{0}}
\def\vnabla{\boldsymbol{\nabla}}

\def\vGamma{\boldsymbol{\Gamma}}

\def\vPi{\boldsymbol{\Pi}}
\def\vphi{\boldsymbol{\phi}}

\def\vtau{\boldsymbol{\tau}}

\def\vXi{\boldsymbol{\Xi}}

\def\vOmega{\boldsymbol{\Omega}}
\def\L{\langle}
\def\R{\rangle}
\newlength{\FigureHeight}
\newlength{\FigureHeightHalf}

\numberwithin{equation}{section}
\titleformat{\section}
{\large\bfseries}{\thetitle.}{0.5em}{}
\titleformat{\subsection}
{\normalfont\itshape\filcenter}{\normalfont\thetitle.}{0.5em}{}
\begin{document}

\title{\vspace{0.0em} Covariance and objectivity in mechanics and turbulence\\
{\Large\emph{A revisiting of definitions and applications}}}
\author{Michael Frewer\thanks{Email address for correspondence:
frewer.science@gmail.com}\\
\small Tr\"ubnerstr.$\,$42, 69121 Heidelberg, Germany}

\date{{\small\today}}
\clearpage \maketitle \thispagestyle{empty}

\vspace{0.0em}\begin{abstract}

\noindent Form-invariance (covariance) and frame-indifference
(objectivity) are two notions in classical continuum mechanics
which have attracted much attention and controversy over the past
decades. Particularly in turbulence modelling it seems that there
still is a need for clarification. The aim and purpose of this
study is fourfold: \emph{(i)} To achieve consensus in general on
definitions and principles when trying to establish an invariant
theory for modelling constitutive structures and dynamic processes
in mechanics, where special focus is put on the principle of
Material Frame-Indifference (MFI), a general principle to model
the response of a material in solid and fluid mechanics, stating
that only frame-indifferent (objective) terms should enter the
constitutive equations. \emph{(ii)} To show that in constitutive
modelling MFI can only be regarded as an approximation that needs
to be reduced to a weaker statement when trying to advance it to
an axiom of nature. \emph{(iii)} To convince that in dynamical
modelling, as in turbulence, MFI may not be utilized as a
modelling guideline, not even in an approximative sense. Instead,
its reduced form has to be supplemented by a second, independent
axiom that includes the microscopic (fluctuating) description of
the dynamical processes. Concerning Navier-Stokes turbulence, the
axiom of Turbulent Frame-Indifference (TFI) is stated in which
turbulence has to be modelled consistently with the invariant
properties of the deterministic Navier-Stokes equations, and
finally \emph{(iv)} to propose a novel invariant modelling ansatz
both for constitutive and dynamical modelling that allows to
include the (mean) velocity field as an own independent modelling
variable, however, not in an absolute but only in the relative
sense as a velocity difference; a result that would systematically
improve current modelling procedures in extended thermodynamics
and turbulence theory to be more consistent with physical
observations.

\vspace{0.5em}\noindent{\footnotesize{\bf Keywords:} {\it Tensors,
Form-invariance, Covariance, Frame-indifference, Objectivity,
Symmetry,\linebreak Inertial and Non-inertial systems, Galilean
and Euclidean transformations, Mach's Principle,\linebreak
Constitutive equations, Acceleration-sensitive
materials, Extended Thermodynamics, Turbulence}}
\end{abstract}

\newpage
\thispagestyle{empty}
\tableofcontents
\newpage

\pagenumbering{arabic}\setcounter{page}{1}

\newgeometry{left=2.5cm,right=2.5cm,top=2.5cm,bottom=2.20cm,headsep=1em}

\section{Introduction\label{Sec1}}

The still ongoing confusion between {\it form-invariance} and {\it
frame-indifference} is the root for many wrong conclusions when
trying to model constitutive structures and dynamical processes in
continuum mechanics. The aim of this study is to clarify the
essential difference between these two concepts as precise as
possible. As a result a universal axiom for invariant modelling is
formulated that clearly separates the statements on
form-invariance and frame-indifference, allowing thus to see that
the velocity field itself expressed as a velocity difference is
indeed a valid modelling variable, both for constitutive and
dynamical modelling, and, hence, that it would be inappropriate to
exclude it from the modelling process as standardly practiced in
the community of classical continuum mechanics. For turbulence
modelling, however, an additional, independent axiom is needed to
ensure full consistency with the known invariant properties of the
deterministic Navier-Stokes equations. Worthwhile to mention here
is that the following mathematical development could entirely be
formulated in a 3D context, without changing the geometrical
description to a 4D framework, as it had been done in
\cite{Frewer09.1} --- nevertheless an adequate 4D-embedding in
classical mechanics is still the most natural framework to discuss
form-invariance and frame-indifference and its consequences for
constitutive modelling in general
\citep{Havas64,Matolcsi06,Matolcsi07,Rouhaud13,Panicaud14,Panicaud16,Van15}
and turbulence modelling in particular
\citep{Frewer09.2,Frewer09.3}.

Before the difference between form-invariance and
frame-indifference can be made clear, it is necessary to have a
clear picture of a what a `tensor' is. First of all, it is not
correct to call every object that carries indices automatically a
tensor. A tensor is more than just an array or an ordered
collection of numbers or components. It has to satisfy a
well-defined transformation law under a certain coordinate
transformation, namely that if any ordered collection of numbers
or components make up a tensor relative to a specified coordinate
transformation then this collection has to transform homogeneously
and multilinearly (locally in each space-time~point). Exactly this
behavior defines the tensor-concept as it was originally
introduced by Voigt, Ricci and Levi-Civita in the end of the 19th
century, and not elsewise. For example, a particular feature of a
tensor is that {\it if} it is zero in one coordinate frame, {\it
then} it is zero in all frames relative to the coordinate
transformations considered.\footnote[2]{Since the tensor concept
is defined on coordinate transformations that have to be
invertible (non-zero Jacobian), this particular feature of a
tensor can also be stated oppositely: {\it If} a tensor is
non-zero in one coordinate frame, {\it then} it is non-zero in all
frames relative to the coordinate transformations considered. Of
course, in general this feature is only to be understood in a
local sense around each point in the manifold separately and not
globally for all points simultaneously
\citep{Schroedinger50}.\label{N1}}

Now, for an array of mathematical objects which transforms as a
tensor, one says that this array transforms form-invariantly. But
form-invariance, i.e., the tensor property of this array does not
imply frame-indifference for this array. In other words, although
each observer will `see' a tensor, or a tensor relation, in the
same structural form, its numerical evaluation, however, will be
different for each observer due to their frame used. Hence
frame-indifference under a specified coordinate transformation is
extraordinarily much more than just form-invariance; it is a much
stronger statement (see also e.g. \cite{Sadiki96}).\linebreak To
be more precise: {\it Form-invariance} applies if a mathematical
object transforms homogeneously and multilinearly under a given
coordinate transformation, while {\it frame-indifference} applies
if this coordinate transformation additionally also acts as a
symmetry transformation on that object. Therefore
frame-indifference must be proven independently and cannot be read
off from any form-invariant expression, or, stated differently:
Frame-indifference implies form-invariance under pure space-time
coordinate transformations, but not vice versa.

Before continuing the discussion on the physical relevance of
these two invariant formula-\linebreak tions, some remarks ought
to be made about the terminology standardly used in the
literature. In the physics community the concept of
`form-invariance' is also termed as `covariance', stemming from
the notion of working with relativistic theories. Unfortunately in
the engineering\linebreak community,\hfill however,\hfill it\hfill
is\hfill frequently\hfill observed\hfill that\hfill
`form-invariance'\hfill is\hfill misleadingly\hfill termed\hfill
as

\newgeometry{left=2.5cm,right=2.5cm,top=2.5cm,bottom=2.4cm,headsep=1em}

\noindent `frame-indifference' (see e.g.
\cite{Truesdell65,Speziale98,Dafalias11,Kirwan16}), thus leading
to great confusion in the literature between the form-invariance
of an equation and its frame-independence. In this regard it is
important to note that the word\linebreak `objectivity' will be
used herein in the following as a synonym for
`frame-indifference', unlike some authors in classical continuum
mechanics (see e.g. \cite{Hutter04,Ariki15,Liu16}) who take it as
synonym for `form-invariance' due to its relation to geometric
invariance (see Appendix~\ref{SecA}), which, in my opinion, does
not properly match as a synonym. The word `objective' should be
used as it is defined in the general linguistic sense, as
something opposing a subjective impression or a measurement.
Frame-indifference is of that category, while form-invariance is
not.

For example, transforming between inertial coordinate systems
within a physically closed environment is a frame-indifferent
(objective) process,\footnote[2]{The coordinate transformations
connecting inertial frames are the Lorentz transformations, or in
the classical limit, the Galilei transformations which rest on
four fundamental symmetries characterizing the underlying
space-time structure: the homogeneity of time, the homogeneity of
space, the isotropy of space and the relativity of space-time. A
detailed explanation of these symmetries can be found, e.g., in
Sec.$\,$7 in \cite{Frewer09.1}.} experimentally first realized by
Galilei that one inertial system can not be distinguished from any
other one; no matter which physical experiment is conducted, they
will always give the same exact (numerical) results in each and
every chosen inertial system --- also known as the equivalence
principle of inertial reference frames, which constitutes a
fundamental principle of nature for all physics, without any
exceptions. In clear contrast when transforming between inertial
and non-inertial or between different non-inertial coordinate
systems, which always, at least in 4D, is only a form-invariant
process,\footnote[3]{Within a classical 3D formulation, the
general form-invariance of physics between arbitrary space-time
coordinate transformations cannot be naturally seen; for that the
theory has to be geometrically reformulated into a true 4D
framework, which can always achieved without changing the physical
content of theory (see also Kretschmann's objection in Section
\ref{Sec4.1}). For example, in
\cite{Frewer09.1,Frewer09.2,Frewer09.3} it has been demonstrated
how the theory of the classical Navier-Stokes equations can be put
into a general form-invariant (covariant) form for arbitrary
space-time transformations without changing the physical content
of these equations, by just reformulating them onto a true 4D
space-time manifold (either flat or curved) within Newtonian
physics.} but definitely not a frame-indifferent (objective) one,
since for each observer it is always possible to distinguish one
non-inertial system from any other one by performing an
appropriate physical experiment (e.g. the Foucault pendulum to
demonstrate the rotation of the earth within a closed room without
any sensuous contact to the outside world). Hence an equivalence
principle for non-inertial reference frames does not exist as it
does for inertial frames.\footnote[4]{Important to note here is
that Einstein's theory of general relativity is not a theory on
general frame-independency. In general relativity it is still
possible for an observer to distinguish between inertial and
non-inertial systems. For example, a free falling observer as an
astronaut will notice the effect of his acceleration induced by
gravity if he inspects larger space-time regions than his
immediate surrounding (e.g. for him the planetary orbits are no
straight lines). In this sense the name `general relativity
theory' can be criticized, since this theory presents no true
generalization of the relativity or equivalence principle for
inertial reference frames towards a relativity or equivalence
principle for non-inertial systems. Essentially, general
relativity is `only' a theory on connecting inertial frames
locally, in contrast to special relativity which is a global
theory on inertial~frames.}

\section{The mathematical description of form-invariance vs. frame-indifference\label{Sec2}}

For the defining coordinate transformation, let's consider for
simplicity in the following only a change in spatial coordinates
given by an orientated uniform (time-dependent) rotation
\begin{equation}
\tilde{\vx}=\vQ  \vx, \;\text{with}\;\,
\vQ\vQ^T=\boldsymbol{1},\;\, \text{det}(\vQ)=1, \;\text{and
constant spin}\;\,  \vOmega := \vQ\dot{\vQ}^T, \;\,
\dot{\vOmega}=\boldsymbol{0},\label{161016:1904}
\end{equation}
which can be interpreted either as a passive or as an active
rotation \citep{Frewer09.1}. Hence the bold-face notation used in
this study is contextual, depending on whether \eqref{161016:1904}
is interpreted passively (change of frame) or actively (change of
state): For a passive interpretation the bold-face notation of the
coordinate $\vx$ has to be seen only as a compact short-hand
notation to symbolize the collection of all components
$\vx^T=(x^1,x^2,x^3)$, thus being equivalent to the index
notation; it does {\it not} symbolize the geometrical,
rotationally invariant coordinate vector $\vx$ itself,

\newgeometry{left=2.5cm,right=2.5cm,top=2.5cm,bottom=2.3cm,headsep=1em}

\noindent which, as an element of vector space with a specifically
chosen basis $\vg_i$, would be represented\linebreak as
$\vx=q^i\vg_i$ being invariant under a rotated change of frame
(passive coordinate transformation):\linebreak
$\vx=q^i\vg_i=\tilde{q}^i\tilde{\vg}_i$. Hence, under a passive
transformation, $\tilde{\vx}$ and $\vx$ in \eqref{161016:1904} do
not represent invariant geometrical objects but only the
(frame-dependent) set of components of the point vector in each
frame, i.e., under a passive transformation, $\tilde{\vx}=\vQ\vx$
equivalently stands~for
\begin{equation}
\tilde{\vx}=\vQ\vx\;\;\;\xrightarrow{\text{passive transf.}}\;\;\;
\tilde{x}^i=(\vQ)^i_{\,\, j} x^j.\label{161024:0038}
\end{equation}
However, for an active interpretation of transformation
\eqref{161016:1904}, the bold-face notation of $\vx$ has to be
seen as an element of vector space $\vx=x^i\vg_i$ being mapped by
a linear mapping $\vQ$ into itself, $\vx\rightarrow
\tilde{\vx}=\vQ\vx$, where the components of the new (transformed)
vector $\tilde{\vx}$ in the old basis $\vg_i$ are given as
\citep{Sexl01}:
$\tilde{\vx}=\vQ(x^i\vg_i)=x^i\vQ(\vg_i)=x^i(\vQ)_i^{\,\,
j}\vg_j=\tilde{x}^j\vg_j$, with $\tilde{x}^j=(\vQ^T)^j_{\,\,
i}x^i$, i.e., under an active transformation, $\tilde{\vx}=\vQ\vx$
equivalently stands for
\begin{equation}
\tilde{\vx}=\vQ\vx\;\;\;\xrightarrow{\text{active transf.}}\;\;\;
\tilde{x}^i=(\vQ^T)^i_{\,\, j} x^j.\label{161024:0039}
\end{equation}
Hence, the actively transformed components \eqref{161024:0039}
relative to one frame (one observer, two states) transform
contragrediently to the passively transformed ones
\eqref{161024:0038} between two frames (two observers, one state).
Of course, with respect to the new basis $\tilde{\vg}_i$ the
actively rotated vector $\tilde{\vx}$ \eqref{161016:1904} has the
same components as $\vx$  has with respect to the original one
$\vg_i$.

As already said, since the defining transformation
$\tilde{\vx}=\vQ\vx$ in \eqref{161016:1904} can be interpreted
either passively or actively in the straightforward way as shown
above, it is not necessary to always explicitly distinguish
between these two cases. Hence, as long as no ambiguity arises, I
will thus consider $\tilde{\vx}=\vQ\vx$ \eqref{161016:1904} in the
following as a generic transformation between two systems
$\tilde{\vx}$ and~$\vx$, without explicitly specifying each time
whether these two systems correspond to two different observers
looking at one state or to only one observer looking at two
different states.

\subsection{The tensor field\label{Sec2.1}}

By considering first an arbitrary single-valued function
$\phi(\vx)$, it will transform under \eqref{161016:1904} as
\begin{equation}
\phi(\vx) = \phi(\vQ^T\tilde{\vx}) =:
\tilde{\phi}(\tilde{\vx}),\label{161106:1017}
\end{equation}
where for notational abbreviation an additional symbol
$\tilde{\phi}$ has been introduced to show that the functional
dependence on the new coordinates will in general be different
than the functional dependence on the old coordinates. This
relation $\phi(\vx)=\tilde{\phi}(\tilde{\vx})$, however, only
tells us that the single-valued function $\phi$ transforms as a
scalar, or as a tensor of rank 0, i.e. form-invariantly or
covariantly. It does not tell us that $\phi$ transforms
frame-indifferently or objectively. In order to transform
frame-indifferently, the function itself must show in addition to
this scalar relation an invariant structure, namely such that if
the new coordinates are inserted into $\phi$, the functional
dependence will stay unchanged
\begin{equation}
\tilde{\phi}(\tilde{\vx})=\phi(\vx) = \phi(\vQ^T\tilde{\vx}) =
\phi(\tilde{\vx}), \quad\text{i.e.}\;\;
\tilde{\phi}(\cdot)=\phi(\cdot).\label{161106:1009}
\end{equation}
Such a coordinate transformation \eqref{161016:1904} is then
called a symmetry transformation for $\phi(\vx)$, while
$\phi(\vx)$ in this specific case is called an isotropic
(rotationally frame-indifferent) function, an invariant function
of the restricted~form
\begin{equation}
\phi(\vx)=\phi(\lVert\vx\rVert),\label{161106:1013}
\end{equation}
 where
$\lVert\vx\rVert=\sqrt{\vx^T\cdot\vx\,}$ is the Euclidean norm
(magnitude) of $\vx$. Indeed, restriction \eqref{161106:1013}
globally satisfies the frame-indifference condition
\eqref{161106:1009} for the specific coordinate transformation
\eqref{161016:1904}.

A tensor in the next higher rank can be constructed by taking the
gradient of the scalar field $\vG(\vx):=\vnabla\phi(\vx)$. Using
the above scalar transformation law for $\phi$ \eqref{161106:1017}
and the chain rule for the derivative, the gradient of a scalar
field transforms as
\begin{equation}
\vG(\vx) = \vnabla\phi(\vx) = \vnabla\tilde{\phi}(\tilde{\vx}) =
\tilde{\vnabla}\frac{\partial\tilde{\vx}}{\partial\vx}
\tilde{\phi}(\tilde{\vx})
=\tilde{\vnabla}\tilde{\phi}(\tilde{\vx})\vQ=:
\tilde{\vG}(\tilde{\vx})\vQ.
\end{equation}

\restoregeometry

\noindent By inverting this relation we get the transformation law
for the gradient of an arbitrary scalar field under a uniform
rotation:
\begin{equation}
\tilde{\vG}(\tilde{\vx})=\vG(\vx)\vQ^T.\label{161013:1108}
\end{equation}
As before, this relation only tells us that the gradient of a
scalar field transforms as a tensor of rank 1, or more precisely
as a covariant tensor of rank 1 due to the appearance of the
transpose (inverse) transformation matrix.\footnote[2]{In this
regard note that the homogeneous and linear transformation for the
coordinates \eqref{161016:1904} defines the contravariant tensor
relation of rank 1, opposed to the covariant relation
\eqref{161013:1108}. However, for all more general spatial
coordinate transformations $\tilde{\vx}=\tilde{\vx}(\vx,t)\neq
\vA(t)\cdot\vx$, which cannot be represented as a homogeneous
linear transformation anymore, the (absolute) coordinates
themselves no longer form a tensor, only the infinitesimal
coordinate differentials $d\tilde{\vx}$ then form a contravariant
tensor of rank 1; but in a pure 3D formulation only if the spatial
coordinate transformations are time-independent, i.e.
$\tilde{\vx}=\tilde{\vx}(\vx)$, since only then the tensor
property is given $d\tilde{\vx}=\partial
\tilde{\vx}/\partial\vx\cdot d\vx$, otherwise we would have the
inhomogeneous relation
$d\tilde{\vx}=\partial\tilde{\vx}/\partial\vx\cdot
d\vx+\partial\tilde{\vx}/\partial t\cdot dt$, which only in a true
4D formulation turns into a tensor relation (see Appendix
\ref{SecA}). In this regard it is also worthwhile to study the
excellent book of \cite{Schroedinger50}, who intriguingly
demonstrates that besides the tensor concept also other concepts
as densities are necessary in order to construct a complete
form-invariant (local) theory.} But this relation does not imply
that $\vG$ transforms frame-indifferently or objectively. Only if
$\vG$ has the following invariant property $\tilde{\vG}(\cdot)=
\vG(\cdot)$ when inserting the new coordinates, then $\vG$ is said
to transform frame-indifferently. But this property can only be
achieved if $\vG$ behaves as follows
\begin{equation}
\vG(\vx) =
\vG(\vQ^T\tilde{\vx})=\vG(\tilde{\vx})\vQ,\quad\text{i.e.}\;\;
\vG(\tilde{\vx})=\vG(\vx)\vQ^T,\label{161012:0911}
\end{equation}
because only then the frame-indifferent transformation rule is
obtained:
\begin{equation}
\tilde{\vG}(\tilde{\vx})= \vG(\vx)\vQ^T=\vG(\vQ^T\tilde{\vx})\vQ^T
=\vG(\tilde{\vx})\vQ\vQ^T= \vG(\tilde{\vx}),\quad\text{i.e.}\;\;
\tilde{\vG}(\cdot)= \vG(\cdot).
\end{equation}
A typical example for an isotropic covariant tensor of rank 1 is
obviously
\begin{equation}
\vG(\vx)=\nabla\phi(\lVert\vx\rVert)=:\frac{\partial}{\partial
\vx} \phi(\lVert\vx\rVert),\label{161012:0910}
\end{equation}
where $\phi$ is again some arbitrary scalar function of the
Euclidean norm $\lVert\vx\rVert=\sqrt{\vx^T\cdot\vx\,}$. Indeed,
structure \eqref{161012:0910} behaves as \eqref{161012:0911}:
\begin{equation}
\vG(\tilde{\vx})=\frac{\partial}{\partial
\tilde{\vx}}\phi(\lVert\tilde{\vx}\rVert)=\frac{\partial}{\partial\vx}
\frac{\partial\vx}{\partial\tilde{\vx}}\phi(\lVert\vQ\vx\rVert)=
\frac{\partial}{\partial
\vx}\phi(\lVert\vx\rVert)\vQ^T=\vG(\vx)\vQ^T.
\end{equation}
Again, it is important to observe the small but decisive
difference between the condition of form-invariance given in
\eqref{161013:1108} and the condition of frame-indifference given
in \eqref{161012:0911}: In relation \eqref{161013:1108} we are
dealing with two different tensor-functions $\tilde{\vG}$ and
$\vG$, while relation \eqref{161012:0911} refers to only one
tensor-function $\vG$.

The above construction procedure can now be straightforwardly
extended to arbitrary order in the rank and type of a tensor. For
example, for a tensor field $\vC=\vC(\vx)$ of rank 2 we can have
the following form-invariant transformation relations
\begin{equation}
\tilde{\vC}(\tilde{\vx})=\vQ\,\vC(\vx)\,\vQ^T,\quad\;\;
\tilde{\vC}(\tilde{\vx})=(\vQ^{-1})^T\vC(\vx)\,\vQ^{-1},\quad\;\;
\tilde{\vC}(\tilde{\vx})=\vQ\,\vC(\vx)\,\vQ^{-1},
\end{equation}
depending on whether the tensor is contravariant $(\vC)^{ij}$,
covariant $(\vC)_{ij}$, or of mixed form $(\vC)^i_{\,\; j}$,
respectively. However, since the considered transformation matrix
$\vQ$ is orthogonal, i.e. since $\vQ^{-1}=\vQ^T$, the above
contravariant and covariant tensors all transform in the same way
for the specifically considered case of uniform rotations,
namely~as
\begin{equation}
\tilde{\vC}(\tilde{\vx})=\vQ\,\vC(\vx)\,\vQ^T,\label{161022:1806}
\end{equation}
but careful, this does not mean that the contravariant and
covariant tensors themselves coincide. The universal relation
\eqref{161022:1806} only states that the components of
contravariant and covariant tensors transform in the same way
under orthogonal transformations, and not that they coincide,
since in general $(\vC)^{ij}\neq (\vC)_{ij}\neq (\vC)^i_{\,\; j}$.
These components will coincide only if we would introduce an
orthonormal basis of the vector space for the coordinate system
considered.\footnote[2]{Note that under orientated orthogonal
transformations (rotations) any chosen basis of a vector space,
irrespective of whether being a contravariant, covariant or
orthonormal basis, remains class invariant.}

That $\vC$ transforms form-invariantly as a tensor
\eqref{161022:1806} is of course not a guarantee that it also
transforms frame-indifferently or objectively. For that it
additionally needs to transform as
\begin{equation}
\vC(\tilde{\vx})=\vQ\vC(\vx)\vQ^T,
\end{equation}
because only then the frame-indifferent transformation rule is
again obtained:
\begin{equation}
\tilde{\vC}(\tilde{\vx})=\vQ\vC(\vx)\vQ^T=\vQ\vQ^T\vC(\tilde{\vx})\vQ\vQ^T=\vC(\tilde{\vx}),
\quad\text{i.e.}\;\; \tilde{\vC}(\cdot)= \vC(\cdot).
\end{equation}
A typical example for a frame-indifferent (objective) second rank
tensor under uniform rotations is the isotropic tensor
\begin{equation}
\vC(\vx)=\phi(\lVert\vx\rVert)\,\vx\otimes\vx+\psi(\lVert\vx\rVert)\,\boldsymbol{1},
\end{equation}
where $\phi$ and $\psi$ are two arbitrary scalar functions of the
spatial Euclidean norm of coordinate $\vx$.

\subsection{The composite tensor field\label{Sec2.2}}

The tensor relations given above refer directly to the coordinates
to be transformed. The question now is: How will these relations
change if we consider nested tensor functions, i.e. composite
tensor functions? Imagine we would have the following scalar
function $\phi(\vf(\vx))$, where for simplicity we will assume
that the scalar function $\phi$ is composed of a contravariant
tensor function $\vf$ of rank 1, i.e., that it transforms as
$\tilde{\vf}(\tilde{\vx})=\vQ\vf(\vx)$. This composed scalar
function will then transform as
\begin{equation}
\phi(\vf(\vx))=\phi(\vQ^T\tilde{\vf}(\tilde{\vx})):=\tilde{\phi}(\tilde{\vf}(\tilde{\vx})),
\label{161106:1412}
\end{equation}
displaying a relation, which formally, if $\vf(\vx)$ is known or
specified, can be written also as
\begin{equation}
\phi^\star(\vx)=\tilde{\phi}^\star(\tilde{\vx}),\label{161106:1413}
\end{equation}
where $\phi^\star$ denotes the function constructed from
evaluating the composite function $\phi(\vf(\vx))$ relative to the
coordinates $\vx$, and likewise the notation $\tilde{\phi}^\star$
for evaluating the transformed composite function, i.e.,
$\phi(\vf(\vx))\equiv\phi^\star(\vx)$ and
$\tilde{\phi}(\tilde{\vf}(\tilde{\vx}))\equiv\tilde{\phi}^\star(\tilde{\vx})$.
Obviously, as already assumed, such a construction is only
possible if the inner function $\vf$ is explicitly known or
specified. Nevertheless, the latter representation
\eqref{161106:1413} refers {\it directly} to the transformed
coordinates, whereas the former representation \eqref{161106:1412}
only refers {\it indirectly} via the inner tensor function $\vf$
to the transformed coordinates. Hence, frame-indifference for this
scalar field $\phi$ can occur in two variants:
\begin{align}
& \text{Explicit frame-indifference, if
$\phi(\cdot)=\tilde{\phi}(\cdot)$},\label{161013:1111}\\[0.5em]
\text{{\it or},}\quad & \text{Full frame-indifference, if
$\phi^\star(\cdot) =
\tilde{\phi}^\star(\cdot)$}.\label{161013:1112}
\end{align}
It is intuitively clear that full frame-indifference
\eqref{161013:1112} is a stronger restriction on $\phi$ than
explicit frame-indifference \eqref{161013:1111}. Full
frame-indifference of $\phi(\vf(\vx))$ can be seen as a
superposition of frame-indifference originating from the outer
dependence $\vf$ (explicit frame-indifference) and the inner
nested dependence $\vx$ (implicit frame-indifference). To
illustrate this difference of restriction explicitly, let's
consider, for example, the following simple
specification$\,$\footnote[3]{Note that the explicit structure of
transformed function $\tilde{\vf}(\tilde{\vx})$ is determined from
the untransformed function $\vf(\vx)$ \eqref{161013:1113} via its
defined tensor transformation rule as
$\tilde{\vf}(\tilde{\vx})\equiv\vQ\vf(\vQ^T\tilde{\vx})$, which
results to: $\tilde{f}^i(\tilde{\vx})=(\vQ)^i_{\;\,
1}(\vQ^T\tilde{\vx})^1$.}
\begin{equation}
\phi(\vf(\vx))=\lVert\vf(\vx)\rVert, \quad \text{with}\;\;
\vf(\vx)^T=(x^1,0,0),\;\;\text{i.e.}\;\;
f^i(\vx)=\delta^i_1\delta^1_j x^j,\label{161013:1113}
\end{equation}
which, as a result, induces the functional relation directly on
the coordinates as
\begin{equation}
\phi(\vf(\vx))=\lVert\vf(\vx)\rVert=\lvert x^1\rvert,
\;\;\text{i.e.}\;\; \phi^\star(\vx)=\lvert x^1\rvert.
\end{equation}
Obviously, the scalar function $\phi(\vf(\vx))$, as specified in
\eqref{161013:1113}, only shows the weaker explicit
frame-indifference \eqref{161013:1111}, since
\begin{equation}
\tilde{\phi}(\tilde{\vf}(\tilde{\vx}))=\phi(\vf(\vx))=\phi(\vQ^T\tilde{\vf}(\tilde{\vx}))
=\lVert\vQ^T\tilde{\vf}(\tilde{\vx})\rVert
=\lVert\tilde{\vf}(\tilde{\vx})\rVert=\phi(\tilde{\vf}(\tilde{\vx}))
,\;\;\text{i.e.}\;\; \phi(\cdot)=\tilde{\phi}(\cdot),
\end{equation}
and {\it not} the stronger full frame-indifference
\eqref{161013:1112}, since$\,$\footnote[2]{Full frame-indifference
can only be obtained if we would specify the contravariant vector
field $\vf(\vx)$ frame-indifferently, e.g., as the isotropic
function $\vf(\vx)=\psi(\lVert\vx\rVert)\,\vx$, where $\psi$ is
some arbitrary scalar function, and where the direct coordinate
dependence of $\phi$ would then read as
$\phi^\star(\vx)=\psi(\lVert\vx\rVert)\,\lVert\vx\rVert$, which
now, obviously, is fully frame-indifferent, since
$\phi^\star(\cdot)\neq\tilde{\phi}^\star(\cdot)$.}
\begin{align*}
\tilde{\phi}^\star(\tilde{\vx})=\tilde{\phi}(\tilde{\vf}(\tilde{\vx}))=\phi(\vf(\vx))&=
\phi^\star(\vx)\\
&=\phi^\star(\vQ^T\tilde{\vx})=\lvert(\vQ^T)^1_{\;\, j}\,
\tilde{x}^j\rvert\,\neq\,
\lvert\tilde{x}^1\rvert=\phi^\star(\tilde{\vx}),\;\;\text{i.e.}\;\;
\phi^\star(\cdot)\neq\tilde{\phi}^\star(\cdot).
\end{align*}
However, as already said before, the representation
\eqref{161106:1413} and its condition for frame-indifference
\eqref{161013:1112} is only then of a concern if the inner
function $\vf$ is known. Usually this is not the case, since
normally such functions $\vf(\vx)$ represent field variables
satisfying certain differential or integral equations which first
have to be solved in order to get hold of the explicit functional
structure of $\vf(\vx)$. Henceforth I will not consider the case
of full frame-indifference \eqref{161013:1112} any further, that
is, throughout this study the notion `frame-indifference' for
composite tensor fields  will only be understood as `explicit
frame-indifference' as formulated in \eqref{161013:1111}.

Surely, the specific dependence $\phi=\phi(\vf(\vx))$ considered
so far is not the most general case for a composite scalar
function. The general case would be
\begin{equation}
\phi=\phi(\vT_1(\vx),\dotsc,\vT_n(\vx)\,; \vx),\label{161006:1610}
\end{equation}
depending explicitly on the coordinates $\vx$ and implicitly on
$n$ different field functions $\vT_n(\vx)$ of any type, not
necessarily tensors. In general these functions are unknown system
variables satisfying certain balance equations, but, of course,
with a known transformation behavior $\vT_n(\vx)\rightarrow
\tilde{\vT}_n(\tilde{\vx})$, $\forall n$, otherwise an {a priori}
invariance analysis on \eqref{161006:1610} would not be~possible.
Form-invariance of \eqref{161006:1610} is then given if $\phi$
transforms as a true scalar, namely as
\begin{equation}
\phi(\vT_1(\vx),\dotsc,\vT_n(\vx)\,;
\vx)=\tilde{\phi}(\tilde{\vT}_1(\tilde{\vx}),\dotsc,\tilde{\vT}_n(\tilde{\vx})\,;
\tilde{\vx}),\label{161106:1720}
\end{equation}
and frame-indifference in the sense as \eqref{161013:1111}, if
$\phi$ additionally behaves as
\begin{equation}
\phi(\tilde{\vT}_1(\tilde{\vx}),\dotsc,\tilde{\vT}_n(\tilde{\vx})\,;
\tilde{\vx})
=\tilde{\phi}(\tilde{\vT}_1(\tilde{\vx}),\dotsc,\tilde{\vT}_n(\tilde{\vx})\,;
\tilde{\vx}).
\end{equation}
The concept of measuring or constructing explicit
frame-indifference as formulated in \eqref{161013:1111} is
conferrable to any composite tensor of any rank. For example, when
looking at the transformation law for the tensor composition
$\vG(\vT_1(\vx),\dotsc,\vT_n(\vx)\,; \vx)$ of rank 1, where $\vG$
is the gradient of the scalar field \eqref{161006:1610}
\begin{equation}
\vG(\vT_1(\vx),\dotsc,\vT_n(\vx)\,; \vx) =
\vnabla\phi(\vT_1(\vx),\dotsc,\vT_n(\vx)\,; \vx),
\end{equation}
which transforms form-invariantly
\begin{equation}
\tilde{\vG}(\tilde{\vT}_1(\tilde{\vx}),\dotsc,\tilde{\vT}_n(\tilde{\vx})\,;
\tilde{\vx})=\vG(\vT_1(\vx),\dotsc,\vT_n(\vx)\,;
\vx)\,\vQ^T,\label{161106:1747}
\end{equation}
since, obviously, according to \eqref{161106:1720}, $\vG$
explicitly transforms as
\begin{align}
\vG(\vT_1(\vx),\dotsc,\vT_n(\vx)\,;
\vx)&=\vnabla\phi(\vT_1(\vx),\dotsc,\vT_n(\vx)\,;
\vx)\nonumber\\[0.5em]
&=\vnabla\tilde{\phi}(\tilde{\vT}_1(\tilde{\vx}),\dotsc,\tilde{\vT}_n(\tilde{\vx})\,;
\tilde{\vx})=
\tilde{\vnabla}\frac{\partial\tilde{\vx}}{\partial\vx}\tilde{\phi}
(\tilde{\vT}_1(\tilde{\vx}),\dotsc,\tilde{\vT}_n(\tilde{\vx})\,;
\tilde{\vx})\qquad\nonumber\\[0.5em]
&=\tilde{\vG}(\tilde{\vT}_1(\tilde{\vx}),\dotsc,\tilde{\vT}_n(\tilde{\vx})\,;
\tilde{\vx})\,\vQ,\label{161106:1737}
\end{align}
the property frame-indifference for $\vG$ in the sense
\eqref{161013:1111}, i.e. $\vG(\cdot)=\tilde{\vG}(\cdot)$, is then
given, if, next to the transformation law \eqref{161106:1737},
$\vG$ additionally transforms as
\begin{equation}
\vG(\tilde{\vT}_1(\tilde{\vx}),\dotsc,\tilde{\vT}_n(\tilde{\vx})\,;
\tilde{\vx})=\vG(\vT_1(\vx),\dotsc,\vT_n(\vx)\,;
\vx)\,\vQ^T,\label{161106:1748}
\end{equation}
a rule analogous to \eqref{161012:0911}, because only then the
condition $\vG(\cdot)=\tilde{\vG}(\cdot)$ is guaranteed:
\begin{align}
\tilde{\vG}(\tilde{\vT}_1(\tilde{\vx}),\dotsc,\tilde{\vT}_n(\tilde{\vx})\,;
\tilde{\vx})&\underset{\eqref{161106:1747}}{=}\vG(\vT_1(\vx),\dotsc,\vT_n(\vx)\,;
\vx)\,\vQ^T\nonumber\\[0.5em]
&\underset{\eqref{161106:1748}}{=}\vG(\tilde{\vT}_1(\tilde{\vx}),\dotsc,\tilde{\vT}_n(\tilde{\vx})\,;
\tilde{\vx}),\;\;\text{i.e.}\;\; \tilde{\vG}(\cdot)=\vG(\cdot).
\end{align}
However, in view of the redefinition problems arising in the
specific examples to be introduced and discussed next, it is
already important to note that to measure or construct explicit
frame-indifference for any tensor~$\vC$ as
$\tilde{\vC}(\cdot)=\vC(\cdot)$, only works for absolute tensors,
but not for (redefined) relative tensors. The notion of absolute
and relative tensors is defined in Appendix~\ref{SecB}, while the
criterion to measure frame-indifference (objectivity) for a
relative tensor is defined in Appendix~\ref{SecC}, where it is
made clear that it is necessary to distinguish between relative
and absolute objectivity.

\subsection{The velocity field --- example for a relative,
non-objective tensor\label{Sec2.3}}

A key quantity in continuum mechanics is the velocity field
$\vu=\vu(\vx,t)$, which formally gets its transformation rule by
taking the time derivative of the transformed coordinates
\eqref{161016:1904}
\begin{align}
\dot{\tilde{\vx}}&=\, \vQ\dot{\vx}+\dot{\vQ}\vx\nonumber\\[0.5em]
& = \vQ\dot{\vx} + \dot{\vQ}\vQ^T\tilde{\vx}=\vQ\dot{\vx}
+\vOmega^T\tilde{\vx}=\vQ\dot{\vx} -\vOmega\tilde{\vx}
\quad\Leftrightarrow\quad \tilde{\vu}+\vOmega\tilde{\vx}=\vQ\vu,
\end{align}
which then, when seen as the transition going from the Lagrangian
to the Eulerian description, yields the well-known transformation
rule for the velocity field$\,$\footnote[2]{To simplify notation
in the following, the explicit time dependence for the velocity
field $\vu$ will be suppressed, which generally shows the full
coordinate dependence $\vu=\vu(\vx,t)$.}
\begin{equation}
\tilde{\vu}(\tilde{\vx})+\vOmega\tilde{\vx}=\vQ\vu(\vx).\label{161020:1425}
\end{equation}
It is clear that the inhomogeneous term proportional to the spin
$\vOmega$ destroys the tensor property of the velocity field in
3D. Nevertheless, as shown and discussed in Appendix \ref{SecB},
it is possible within the 3D framework to reformulate
\eqref{161020:1425}  into a form-invariant, but still
frame-dependent (non-objective) tensor~relation
\begin{equation}
\tilde{\vu}_{\tilde{\vOmega}}(\tilde{\vx})=\vQ\vu_{\vOmega}(\vx),
\end{equation}
with $\vu_{\vOmega}(\vx):=\vu(\vx)+\vOmega\vx$ being the new
redefined velocity field relative to the spin term $\vOmega$,
which itself, however, transforms as a non-tensor
\begin{equation}
\tilde{\vOmega}=\vQ\vOmega\vQ^T+\vQ\dot{\vQ}^T.
\end{equation}
As introduced in Appendix \ref{SecB}, the redefined velocity field
$\vu_{\vOmega}(\vx)$ cannot be classified as a true (absolute)
tensor, but only as a relative one, since obviously
$\vu(\vx)\neq\vu_{\vOmega}(\vx)$, for $\vOmega\neq
\boldsymbol{0}$. While a true tensor can either be
frame-indifferent (objective) or frame-dependent (non-objective),
a relative tensor can never be objective in an absolute sense,
since it always, {\it per se}, depends on the reference frame
used, for example, as already seen at the relative velocity
$\vu_{\vOmega}(\vx)=\vu(\vx)+\vOmega\vx$, which explicitly depends
on the spin value $\vOmega$ of the chosen frame. The upcoming
examples, along with Appendix \ref{SecC}, will clarify this issue
even further.

\subsection{The strain rate --- example for an absolute, objective
composite tensor\label{Sec2.4}}

The next example is the strain rate $\vS$, defined as the
following symmetric composite function of the velocity gradient
$\vL(\vx):=\vnabla\otimes\vu(\vx)$:
\begin{equation}
\vS(\vL(\vx))={\textstyle \frac{1}{2}}\left(\vL(\vx)
+\vL(\vx)^T\right),\label{161013:1115}
\end{equation}
which, as is well known, transforms form-invariantly as a tensor
of rank~2 under a uniform rotation:
\begin{equation}
\tilde{\vS}(\tilde{\vL}(\tilde{\vx}))=\vQ\vS(\vL(\vx))\vQ^T.\label{161013:1116}
\end{equation}
In contrast to the velocity gradient $\vL(\vx)$ which itself does
not transform as tensor, due to the appearance of an inhomogeneous
term arising from the non-tensor transformation property of the
velocity field $\vu(\vx)$ \eqref{161020:1425}:
\begin{multline*}
\vL(\vx) = \vnabla\otimes\vu(\vx) =
\tilde{\nabla}\frac{\partial\tilde{\vx}}{\partial\vx} \otimes
\left(\vQ^T\tilde{\vu}(\tilde{\vx})+\dot{\vQ}^T\tilde{\vx}\right)=
\tilde{\nabla}\frac{\partial\tilde{\vx}}{\partial\vx} \otimes
\left(\vQ^T\tilde{\vu}(\tilde{\vx})+\vQ^T\vOmega\tilde{\vx}\right)\\[0.5em]
=\vQ^T\Big(\tilde{\nabla}\otimes\tilde{\vu}(\tilde{\vx})
+\tilde{\nabla}\otimes\vOmega\tilde{\vx}\Big)\frac{\partial\tilde{\vx}}{\partial\vx}
=\vQ^T\tilde{\vL}(\tilde{\vx})\vQ+\vQ^T\vOmega\vQ,
\end{multline*}
that means, we obtain the {\it non}-tensor relation
\begin{equation}
\tilde{\vL}(\tilde{\vx})=\vQ\vL(\vx)\vQ^T-\vOmega\;\neq\;
\vQ\vL(\vx)\vQ^T.\label{161013:1634}
\end{equation}
Hence, since the spin matrix $\vOmega$ is antisymmetric, i.e.
$\vOmega^T=-\vOmega$, we can see the reason why the strain rate
$\vS$ \eqref{161013:1115} transforms as a tensor formulated by
\eqref{161013:1116}, while its argument, the velocity gradient
$\vL$,~not:\footnote[2]{In the formulation of Appendix \ref{SecB},
note that the strain rate $\vS$ obviously transforms as an
absolute tensor, since it need {\it not} to be redefined in order
to obtain a tensor relation for its transformed values, i.e.,
since for $\vOmega\neq\boldsymbol{0}$ it obviously has the
property $\vS=\vS_{\vOmega}$. In contrast of course to its
argument, the velocity gradient~$\vL$, which need to be redefined
as $\vL\rightarrow \vL_{\vOmega}=\vL+\vOmega$ if one wants to turn
\eqref{161013:1634} into a form-invariant tensor relation as
presented in \eqref{161019:0943}. Hence, since
$\vL\neq\vL_{\vOmega}$ for $\vOmega\neq\boldsymbol{0}$, the
velocity gradient thus only transforms as a relative tensor.} Due
to adding $\vL$ and $\vL^T$, the inhomogeneous spin term cancels
$\vOmega+\vOmega^T=\boldsymbol{0}$.\\
Again, as stated several times before for other quantities,
relation \eqref{161013:1116} only tells us that the strain rate
$\vS$ is transforming as a tensor of rank 2; it does not
automatically tell us that~$\vS$ is also transforming
frame-indifferently under uniform rotations. Such a property has
to be analyzed separately. A closer inspection, however, reveals
that the composite structure \eqref{161013:1115} is indeed
frame-indifferent in an absolute sense by showing the
transformation behavior $\tilde{\vS}(\cdot)=\vS(\cdot)$, since
\begin{align}
\tilde{\vS}(\tilde{\vL}(\tilde{\vx}))&=\vQ\vS(\vL(\vx))\vQ^T
=\vQ\left({\textstyle\frac{1}{2}}\big(\vL(\vx)
+\vL(\vx)^T\big)\right)\vQ^T\nonumber\\[0.5em]
&=\vQ\left({\textstyle\frac{1}{2}}\big(\vQ^T\tilde{\vL}(\tilde{\vx})\vQ+\vQ^T\vOmega\vQ
+\vQ^T\tilde{\vL}(\tilde{\vx})^T\vQ+\vQ^T\vOmega^T\vQ\big)\right)\vQ^T
={\textstyle\frac{1}{2}}\big(\tilde{\vL}(\tilde{\vx})
+\tilde{\vL}(\tilde{\vx})^T\big)\nonumber\\[0.5em]
&=\vS(\tilde{\vL}(\tilde{\vx})).
\end{align}

\subsection{Vorticity --- example for a relative, non-objective composite tensor\label{Sec2.5}}

A different example is the vorticity matrix, formulated as the
antisymmetric composite function of the velocity gradient $\vL$
\begin{equation}
\vW(\vL(\vx))={\textstyle \frac{1}{2}}\left(\vL(\vx)
-\vL(\vx)^T\right),\label{161013:1636}
\end{equation}
which, obviously, does not transform as a tensor, since
\begin{align}
\vW(\vL(\vx))&={\textstyle \frac{1}{2}}\left(\vL(\vx)
-\vL(\vx)^T\right)
={\textstyle\frac{1}{2}}\left(\vQ^T\tilde{\vL}(\tilde{\vx})\vQ+\vQ^T\vOmega\vQ
-\vQ^T\tilde{\vL}(\tilde{\vx})^T\vQ-\vQ^T\vOmega^T\vQ\right)\nonumber\\[0.5em]
&=\vQ^T\tilde{\vW}(\tilde{\vL}(\tilde{\vx}))\vQ +\vQ^T\vOmega\vQ,
\end{align}
that means, we obtain the {\it non}-tensor relation
\begin{equation}
\tilde{\vW}(\tilde{\vL}(\tilde{\vx}))=\vQ\vW(\vL(\vx))\vQ^T-\vOmega\;\neq\;
\vQ\vW(\vL(\vx))\vQ^T,
\end{equation}
being identical to the transformation rule of its argument, the
velocity gradient $\vL$ \eqref{161013:1634}. Now, since the
vorticity matrix $\vW$ \eqref{161013:1636} does not transform as a
tensor under uniform rotations, it is clear that it then also
cannot show any frame-indifference: While frame-indiffer\-ence
(the~symmetry property) always implies form-invariance (the tensor
property) under pure space-time coordinate transformations, the
opposite obviously cannot be stated. Said differently, if some
quantity does not transform as a tensor then it also cannot be
frame-indifferent. This can be readily seen by the simple argument
that if the considered quantity is zero in one frame then it is,
due to its non-tensor property, necessarily not zero and thus
different in another (transformed) frame, exhibiting thus a clear
explicit frame-dependence (non-objectivity).

Up to now for the flow quantities mentioned, it is only the strain
rate $\vS$ \eqref{161013:1115} which transforms as an absolute
tensor being additionally also frame-indifferent under uniform
rotations, while the other three quantities, the velocity field
$\vu$, the velocity gradient $\vL$ and the vorticity $\vW$ only
transform as relative tensors, since in each case they need to be
redefined in order to obtain for their transformed values (within
a 3D spatial framework) a form-invariant, but still
frame-dependent (non-objective) relation (see Appendix
\ref{SecB}):
\begin{equation}
\left.
\begin{aligned}
\tilde{\vu}(\tilde{\vx})+\vOmega\tilde{\vx} =
\vQ\vu(\vx)&\;\rightarrow\;
\tilde{\vu}_{\tilde{\vOmega}}(\tilde{\vx})= \vQ\vu_{\vOmega}(\vx),
\\
\tilde{\vL}(\tilde{\vx})+\vOmega = \vQ\vL(\vx)\vQ^T
&\;\rightarrow\; \tilde{\vL}_{\tilde{\vOmega}}(\tilde{\vx})=
\vQ\vL_{\vOmega}(\vx)\vQ^T ,\\
\tilde{\vW}(\tilde{\vL}(\tilde{\vx}))+\vOmega =
\vQ\vW(\vL(\vx))\vQ^T & \;\rightarrow\;
\tilde{\vW}(\tilde{\vL}_{\tilde{\vOmega}}(\tilde{\vx}))=
\vQ\vW(\vL_{\vOmega}(\vx))\vQ^T,
\end{aligned}
~~~~~ \right\}\label{161017:0801}
\end{equation}
satisfying then in each redefined case the multilinear and
homogeneous transformation property of a (3D~spatial) tensor.
Moreover, as shown in Appendix \ref{SecC}, although the relative
vorticity $\vW(\vL_{\vOmega}(\vx))$ is frame-dependent
(non-objective) in the absolute sense, it nevertheless exhibits a
certain relative frame-indifference (relative objectivity) between
different uniform rotating (non-inertial) reference frames.

\subsection{Example for an absolute, non-objective composite tensor\label{Sec2.6}}

For the last example, we want to construct a flow quantity $\vZ$
that transforms as an absolute tensor but that does not show
frame-indifference. Such quantities are easy to construct and
become relevant when opting for explicit (non-inertial)
frame-dependence in modelling relations. For example, let us
consider $\vZ$ as a composite contravariant tensor field of rank 2
of the form $\vZ=\vZ(\vL(\vx))$ with the following specification
for its components
\begin{equation}
Z^{ij}(\vL(\vx))=\delta^i_1\delta^j_1 \Delta
L^1_1,\label{161013:1831}
\end{equation}
where $L^1_1=\partial_1 u^1$ is the $(1,1)$-component of the
velocity gradient matrix $\vL(\vx)=\nabla\otimes\vu(\vx)$, and
$\Delta=\nabla^T\cdot\nabla$ the Laplacian, being, of course as a
scalar, invariant under arbitrary uniform rotations:
$\tilde{\Delta}=\Delta$. That the collection of all components
\eqref{161013:1831} of $\vZ$ transforms as a (contravariant,
second rank) tensor is easily shown, since
\begin{align}
Z^{ij}(\vL(\vx))&=\delta^i_1\delta^j_1 \Delta L^1_1=(\vQ^T\vQ)^i_1
\, (\vQ^T\vQ)^j_1 \,\tilde{\Delta}
\big(\vQ^T\tilde{\vL}\vQ+\vQ^T\vOmega\vQ\big)^1_1\nonumber\\[0.5em]
& = (\vQ^T)^i_{\;\, k}(\vQ)^k_{\;\,\,
1}\,\,\big(\vQ^T\tilde{\Delta}\tilde{\vL}\vQ\big)^1_1\,\,
(\vQ^T)_1^{\;\,\, l}(\vQ)_l^{\;\, j}\nonumber\\[0.5em]
&=: (\vQ^T)^i_{\;\, k}
\tilde{Z}^{kl}(\tilde{\vL}(\tilde{\vx}))(\vQ)_l^{\;\, j},
\end{align}
that means, we get the tensor relation
\begin{equation}
\vZ(\vL(\vx))=\vQ^T \tilde{\vZ}(\tilde{\vL}(\tilde{\vx}))\vQ,
\;\text{or, in inverted form:}\;\;
\tilde{\vZ}(\tilde{\vL}(\tilde{\vx}))=\vQ \vZ(\vL(\vx))\vQ^T,
\end{equation}
where the components of $\vZ$ in the transformed domain are
collectively then given as
\begin{equation}
\tilde{Z}^{ij}(\tilde{\vL}(\tilde{\vx}))=(\vQ)^i_{\;\,\,
1}\,\,\big(\vQ^T\tilde{\Delta}\tilde{\vL}\,\vQ\big)^1_1\,\,
(\vQ^T)_1^{\;\,\, j}.
\end{equation}
Now, although $\vZ$ \eqref{161013:1831} transforms as a tensor in
the absolute sense, it does not transform frame-indifferently,
i.e., $\tilde{\vZ}(\cdot)\neq\vZ(\cdot)$, or, when formulated in
its components, $\tilde{Z}^{ij}(\cdot)\neq Z^{ij}(\cdot)$, since
obviously
\begin{align}
\tilde{Z}^{ij}(\tilde{\vL}(\tilde{\vx}))&=(\vQ)^i_{\;\,
1}\,\big(\vQ^T\tilde{\Delta}\tilde{\vL}\,\vQ\big)^1_1\,
(\vQ^T)_1^{\;\, j}\nonumber\\[0.5em]
&=(\vQ)^i_{\;\, 1}(\vQ^T)^1_{\;\,
k}\,\tilde{\Delta}\tilde{L}^k_l\, (\vQ)^l_{\;\, 1}(\vQ^T)_1^{\;\,
j} \;\neq\; \delta^i_1\delta^j_1 \tilde{\Delta}\tilde{L}^1_1
=Z^{ij}(\tilde{\vL}(\tilde{\vx})).
\end{align}

\section{Galilean form-invariance vs. frame-indifference in a physical application\label{Sec3}}

To see the difference and consequences of form-invariance
(covariance) and frame-indifference (objectivity) within a
physical application as clearly as possible, I will only consider
in this section the theory of classical mechanics for a single
particle, however, which in each step can be easily transferred to
the notions of continuum mechanics.

The equation of motion for a single particle in an inertial system
in 3D is given by Newton's second~law
\begin{equation}
m\ddot{\vx}=\vF,\label{161031:1517}
\end{equation}
where $m$ is the mass and $\vF$ the given force to cause the
acceleration $\ddot{\vx}$ of the particle. The coordinate
transformations connecting all inertial systems in classical
Newtonian physics are the Galilei transformations
\begin{equation}
\vx^\prime=\vR\hspace{0.5mm}\vx+\vv\hspace{0.5mm}t+\vc,\qquad
t^\prime=t+\tau,\label{161031:1518}
\end{equation}
forming mathematically a proper, orthochronous Lie-group with 10
constant parameters:\linebreak the eigenvector (axis of rotation)
of the 3D rotation matrix $\vR$ (having the property
$\vR\vR^T=\boldsymbol{1}$, $\text{det}(\vR)=1$), the velocity
boost $\vv$, the spatial shift $\vc$ and the temporal off-set
$\tau$. Now, transforming Newton's law of motion
\eqref{161031:1517} according to \eqref{161031:1518}, one obtains
the defining result that it is form-invariant in all inertial
systems, since, after transformation, the same form of
\eqref{161031:1517} is again obtained
\begin{equation}
m\ddot{\vx}^\prime=\vF^\prime,\label{161031:1550}
\end{equation}
where $\vF^\prime$ is the same force $\vF$ only expressed in the
new `primed' coordinates related to the components of $\vF$ by
\begin{equation}
\vF^\prime=\vR\hspace{0.5mm}\vF.\label{161031:1701}
\end{equation}
Hence all observers experience the same law of motion
\eqref{161031:1517} only expressed in their
coordinates~\eqref{161031:1550}, that is, $\vF$ and $\vF^\prime$
is the same physical force only in different component
representations belonging to different observers. Obviously this
form-invariance is caused by the fact that the acceleration
$\ddot{\vx}$ in \eqref{161031:1517} transforms as a tensor under
\eqref{161031:1518}
\begin{equation*}
\ddot{\vx}=\frac{d}{dt}\frac{d\vx}{dt}=\frac{d}{dt}
\frac{\frac{\partial \vx}{\partial
\vx^\prime}d\vx^\prime+\frac{\partial \vx}{\partial
t^\prime}dt^\prime} {\frac{\partial t}{\partial
\vx^\prime}d\vx^\prime+\frac{\partial t}{\partial t^\prime}
dt^\prime}=\frac{d}{dt}\left(\frac{\vR^Td\vx^\prime-\vR^T\vv\hspace{0.5mm}
dt^\prime}{dt^\prime}\right)=\frac{d}{dt^\prime}\left(\vR^Td\dot{\vx}^\prime-\vR^T\vv\right)
=\vR^T\ddot{\vx}^\prime,
\end{equation*}
explicitly showing its tensor property $\ddot{\vx}^\prime=\vR
\hspace{0.5mm}\ddot{\vx}$ in transforming linearly and
homogeneously.\footnote[2]{Note that for $\vv\neq\boldsymbol{0}$
and $\vc\neq\boldsymbol{0}$ the spatial coordinates $\vx^\prime$
\eqref{161031:1518} themselves do {\it not} transform
form-invariantly as a tensor; see also Appendix \ref{SecA}.} And
since Newton's law \eqref{161031:1517} forms a mathematical
equation, the proven tensor property of $\ddot{\vx}$ on the
left-hand side of equation \eqref{161031:1517} then defines its
right-side, the force $\vF$, to be a tensor of the same type, too.
Hence the form-invariant tensor relation \eqref{161031:1701},
stating that if the physical force $\vF$ is observed as zero in
one inertial system, it is observed as zero in {\it all} inertial
systems.

However, up to now we have only shown the weaker form-invariance
of Newton's law of motion \eqref{161031:1517} under the Galilei
transformations \eqref{161031:1518}, and not the stronger
frame-indifference. According to the notions and laws of
mechanics, space and time are such that no point and no direction
in (empty) space is distinguished, that time can be chosen
arbitrarily, and that an absolute velocity (except for the speed
of light in vacuum) cannot be defined. In classical Newtonian
mechanics all these facts are expressed by the Galilei
transformations \eqref{161031:1518}, where experiments, first
conducted by Galilei, show that within a physically closed
environment\footnote[3]{Under a `closed system' we only understand
a system in which the sources of all forces are part of the
system, i.e., a system where no external forces occur
\citep{Stephani04}; see also second footnote on
p.$\,$\pageref{N3}. Of course, a closed system is only an
approximation, because, as correctly stated by \cite{Stephani04},
how far do we have to go to get a really closed system? Is our
Galaxy sufficient, or do we have to take the whole universe?} no
inertial system is distinguished from any other one in the sense
that, for {\it any} physical experiment conducted, all inertial
systems give the same exact (numerical) results. This experimental
fact forms the basis for the equivalence principle of inertial
reference frames, stating that the laws of motion for closed
systems are not only form-invariant (covariant), but also
frame-indifferent (objective). In other words, not only do the
laws have the same form in all inertial systems, but also the
processes (induced by these laws) run in the same way in all of
these systems.

Hence, according to this principle the tensorial equation
\eqref{161031:1517} also has to be frame-indifferent under the
Galilei transformations \eqref{161031:1518}. As a consequence, the
force $\vF$ can no longer be treated as a fully arbitrary quantity
anymore, it has to be specified in terms of the basic mechanical
quantities in order to obtain a particular system of differential
equations such that the Galilei transformations
\eqref{161031:1518} are admitted as symmetries. In other words,
the force $\vF$ has to be identified and formulated as a
constitutive relation which has to obey the principle of Galilean
frame-indifference --- similar to the process of modelling
constitutive equations in continuum mechanics according to the
principle of MFI, or in its correctly stated form, according to
r-MFI (see definition and discussion in Section \ref{Sec4.2} \&
\ref{Sec4.3}). In general the force $\vF$ in \eqref{161031:1517}
may explicitly depend on time, space, direction and velocity, as
well as on the position and velocity of the~particle:
\begin{equation}
\vF=\vF(\vx,\dot{\vx},t;\vx^r_0,\vv^r_0,t^r_0),\label{161101:0103}
\end{equation}
where $\vx^r_0$, $\vv^r_0$ and $t_0^r$ are some arbitrary but
fixed reference values of the system, subject to be transformed in
the same way as the position $\vx$, the velocity $\dot{\vx}$, and
the evolution coordinate~$t$ of the particle for some constant
value, respectively.\footnote[4]{To note here is that the
reference values $\vx^r_0$, $\vv^r_0$ and $t_0^r$ do {\it not}
represent the initial position $\vx_0:=\vx(t_0)$ and velocity
$\dot{\vx}_0:=\dot{\vx}(t_0)$ of the particle at the initial time
$t_0$. In other words these reference values are not
characteristics of the particle, but of the system in which this
particle evolves. For example, the absolute point $\vx_0^r$ can be
some fixed reference point in the laboratory, $\vv_0^r$ the
absolute velocity with which the experiment may move or be
superposed with in that laboratory, and $t^r_0$ the absolute time
point in the day or in the week at which the experiment is
conducted.} Since $\vF$~\eqref{161031:1701} transforms as a
(contravariant) tensor field of rank~1, the condition for
frame-indifference $\vF^\prime(\cdot)=\vF(\cdot)$, as it was
correspondingly derived in~\eqref{161012:0911} for a covariant
tensor field, leads then to the following restricting relation for
the~force
\begin{equation}
\vF\big(\vx^\prime,\dot{\vx}^\prime,t^\prime;
\vx_0^{r\prime},\vv_0^{r\prime},t_0^{r\prime}\big)=\vR
\hspace{0.5mm}\vF(\vx,\dot{\vx},t;\vx^r_0,\vv^r_0,t^r_0).\label{161021:2317}
\end{equation}
To solve such a relation in the most general way is part of a
Lie-group classification problem (see e.g
\cite{Bihlo12,Ibragimov16,Kontogiorgis16}). The most obvious
solution of \eqref{161021:2317} is the particular solution
\begin{equation}
\vF(\vx,\dot{\vx},t;\vx^r_0,\vv^r_0,t^r_0)=f_1\cdot
(\vx-\vx^r_0)+f_2\cdot (\dot{\vx}-\vv^r_0),\label{161101:1018}
\end{equation}
where the two $f_i$'s are arbitrary scalar functions of the
general form
\begin{equation}
f_i=f_i(\lVert
\vx-\vx^r_0\rVert,\lVert\dot{\vx}-\vv^r_0\rVert,(\vx-\vx^r_0)^T\cdot
(\dot{\vx}-\vv^r_0),t-t_0^r),
\end{equation}
which ultimately states that for closed systems the laws of nature
do not permit an experimental verification, or a sensible
definition, of an absolute location in space and time, or of an
absolute direction in space, or, particularly in classical
Newtonian mechanics, of an absolute velocity. Only relative
distances, time differences, directional differences and relative
speeds enter the constitutive relation~\eqref{161101:0103}.

Specifying the Galilean frame-indifferent constitutive relation
\eqref{161101:1018} leads to a particular physical problem. For
example, choosing $f_1=-\kappa$ as a constant and $f_2=0$, leads
in 1D to Hooke's~law, turning Newton's law of motion
\eqref{161031:1517} into the problem of the harmonic oscillator
\begin{equation}
m\ddot{x}=-\kappa\, (x-x_0^r),\label{161101:1106}
\end{equation}
where the absolute reference point $x_0^r$ can be for example the
suspension point of the spring. As correctly noted by
\cite{Horzela91}, it is important to stress here the fact that the
individual terms on the right-hand side in \eqref{161101:1106} do
not have the meaning of a force.\footnote[2]{This misleading
interpretation stems from the fact that one is always used to
solve the harmonic oscillator in its non-covariant form, namely as
$m\ddot{x}=-\kappa\,x$, being however only valid in one particular
reference frame in which the center (equilibrium point) of the
oscillator is at rest \citep{Horzela91,Horzela92}.} Only the
difference on the right-hand of \eqref{161101:1106} has a physical
meaning of a force and it is erroneously to think that the
right-hand side is a difference of two forces.

Another example would be to choose $f_1=-mg/\lVert
\vx-\vx^r_0\rVert$, where $g=G\cdot M_{\text{E}}/r_\text{E}^2$ is
the Earth's gravitational acceleration on its surface with $G$ as
the universal gravitational constant and $M_\text{E}$ and
$r_\text{E}$ as the Earth's mass and radius, and $f_2=-a$ as a
constant proportional to the drag coefficient of air, then
Newton's law of motion \eqref{161031:1517} turns into the problem
of gravitational motion (close to the Earth's surface, i.e. where
$\lVert\vx-\vx_0^r\rVert\sim r_\text{E}$) under the influence of
air friction
\begin{equation}
m\ddot{\vx}=-a(\dot{\vx}-\vv_0^r)-mg\frac{\vx-\vx^r_0}{\lVert\vx-\vx^r_0\rVert},
\label{161101:1404}
\end{equation}
where $\vx_0^r$ is the absolute position of the Earth's center and
$\vv_0^r$ the absolute air speed (wind) experienced in the system
when the particle would be at rest \citep{Stephani04}. By
construction, the law of motion \eqref{161101:1404} is
simultaneously form-invariant (covariant) and frame-indifferent
(objective) under Galilei transformations \eqref{161031:1518},
even if, e.g., the wind speed would be experienced\linebreak in
one inertial system as zero, i.e. as $\vv_0^r=\boldsymbol{0}$, the
law \eqref{161101:1404} remains indifferent since in another
inertial system the wind speed may be experienced as non-zero, in
particular as $\vv_0^{r\prime}=\vv$, according
to~\eqref{161031:1518}, such that the velocity difference
$\Delta\dot{\vx}:=\dot{\vx}-\vv_0^r$ always remains
covariant:~$\Delta\dot{\vx}^\prime=\vR\hspace{0.25mm}
\Delta\dot{\vx}$.

Obviously, when transforming to a non-inertial system, say by the
following Euclidean (time-dependent Galilei) transformation
\begin{equation}
\vx^*=\vR(t)\hspace{0.5mm}\vx+\vc(t),\qquad
t^*=t+\tau,\label{161101:1544}
\end{equation}
the form of \eqref{161101:1404} changes and thus is not
frame-indifferent anymore.\footnote[3]{As was already noted
before, under pure coordinate transformations frame-indifference
(objectivity) implies form-invariance (covariance), but not vice
versa. In other words, if a relation is form-invariant then it has
not to be also frame-indifferent, or, stated oppositely, if a
relation is not form-invariant then it cannot be
frame-indifferent.\label{N4}} This change in form then reads
\begin{equation}
m\ddot{\vx}^*=-a(\dot{\vx}^*-\vv_0^{r*})-mg\frac{\vx^*-\vx^{r*}_0}
{\lVert\vx^*-\vx^{r*}_0\rVert}+\vF_\text{Inertial}^*,
\end{equation}
with the fictitious or inertial force given as
\begin{equation}
\vF_\text{Inertial}^*=m\ddot{\vc}-m\vR\ddot{\vR}^T(\vx^*-\vc)-2m\vR\dot{\vR}^T(\dot{\vx}^*-\dot{\vc})
-a\vR\dot{\vR}^T(\vx^*-\vc),\label{161113:1849}
\end{equation}
where here, in the non-inertial case, it is important to know
that, while the velocity of the particle transforms as
\begin{equation}
\dot{\vx}^* =
\vR\hspace{0.5mm}\dot{\vx}+\dot{\vR}\hspace{0.5mm}\vx+\dot{\vc},\label{161113:1824}
\end{equation}
the wind speed $\vv_0^r$ only transforms as it is known in the
inertial case
\begin{equation}
\vv_0^{r*}=\vR\hspace{0.5mm}\vv_0^r+\dot{\vc},\label{161113:1825}
\end{equation}
simply because the wind speed in an inertial system $\vv_0^r$ is a
constant vector, where its components in any new frame are then
just rotated by $\vR$ and translated by $\dot{\vc}$, in contrast
to the particle's velocity~$\dot{\vx}$, which is a differential
quantity and thus needs to be treated differently.

Due to this different non-inertial transformation behavior between
\eqref{161113:1824} and \eqref{161113:1825} we thus obtain the
following important result that should be kept in mind for the
discussion in the next sections: When transforming to a
non-inertial system, it is not only the dynamic part $\ddot{\vx}$
in Newton's second law \eqref{161031:1517} that gives rise to
fictitious or inertial forces, but also the constitutive relation
for the force $\vF$ itself may give frame-dependent contributions
in a very natural way. In the particular case \eqref{161101:1404},
it is the velocity difference $a(\dot{\vx}-\vv_0^r)$ of the
specified constitutive relation~\eqref{161101:1018} that gives
rise to an own contribution $a\vR\dot{\vR}^T(\vx^*-\vc)$ in the
resulting inertial force~\eqref{161113:1849}. In other words,
although the constitutive relation \eqref{161101:1018} itself is
manifestly frame-indifferent for inertial systems by construction,
it nevertheless picks up frame dependent terms when changing to a
non-inertial system.

\section{A comprehensive overview on covariance and objectivity in physics\label{Sec4}}

A view into the literature reveals that when posing
form-invariance (covariance) or frame-indifference (objectivity)
as a principle of nature they both have been praised and
criticized. Let me give a detailed overview on this controversial
situation in each case.

\subsection[The criticism on form-invariance as a principle of nature]
{The criticism on form-invariance as a principle of
nature$\,$\footnote[2]{This section is an excerpt from
Sec.$\,$8.1.2 in \cite{Frewer09.1}.}\label{Sec4.1}}

When Einstein formulated his general theory of relativity in 1915,
he was proud to present a theory that was generally
form-invariant, or as how he first called it, a theory that was
generally covariant: Its equations retained their form under
arbitrary transformations of the space-time coordinate system.
Einstein had the following argument for general covariance: the
physical content of a theory is exhausted by a catalog of events,
which must be preserved under arbitrary coordinate
transformations; all we do in coordinate transformations is just
relabelling the space-time coordinates assigned to each event.
Therefore a physical theory should be generally covariant.
Einstein thus claimed that his general theory of relativity rests
on three physical pillars: on the principle of general covariance,
on the principle of constancy in the speed of light for all local
inertial reference frames, and on the principle of equivalence
between inertial and gravitational mass; only when taking all
three principles together, the theory of general relativity
arrives at a new description of gravitation in terms of a curved
four-dimensional space-time.

Shortly after, \cite{Kretschmann17} pointed out and concurred in
by \cite{Einstein18a} that the principle of general covariance is
fully devoid of any physical content. For, Kretschmann urged, it
is unessential to declare general covariance as a principle of
nature, since any space-time theory whatever can be formulated in
a generally covariant form as long as one is prepared to put
sufficient energy into the task of reformulating it; thus the
theory of general relativity rests on only two pillars, on the
principle of a constant speed of light for all observers in their
immediate neighbourhood and on the mass equivalence principle. In
arriving at general relativity, Einstein had used the tensor
calculus of Ricci and Levi-Civita, where Kretschmann pointed to
this calculus as a mathematical tool that made the task of finding
generally covariant formulations of theories tractable. In his
objection, Kretschmann agreed that the physical content of
space-time theories is exhausted by a catalog of events and that
they should be preserved under any coordinate transformation, but
this, he argued, is no peculiarity of the new gravitational theory
presented by Einstein. For this very reason {\it all} space-time
theories can be given generally covariant formulations. For a
further discussion of Kretschmann's objection, Einstein's response
and of the still active debate that follows, see the articles of
\cite{Norton93,Norton95} and~\cite{Dieks06}.\linebreak
\indent Kretschmann's objection that general covariance is
physically vacuous, in that it does not limit or restrict the
range of acceptable theories, is really a non-trivial objection
when it comes down to constructing and formulating `new' physical
theories, as it was at that time in 1917 for the theory of general
relativity. However, for already existing physical theories his
objection is more or less obvious from a pure theoretical point of
view, since at the end, only a new mathematical representation is
given for the theory. But this does not mean that it is always an
easy task to put any given theory into a generally covariant form;
sometimes it's a challenge to the mathematician's or physicist's
ingenuity.

Even to the classical laws of fluid motion and to the physics of
turbulence modelling, Kretschmann's objection does seem
sustainable. In \cite{Frewer09.1} it is shown how it is possible
to rewrite the Navier-Stokes equations as well as the Reynolds
averaged equations in a {\it general} manifest form-invariant way
without changing the physical content of the theory. For that,
however, one has to change to a four-dimensional space-time
formulation.\footnote[2]{For the specific case of Euclidean
(time-dependent Galilei) coordinate transformations,
\cite{Sadiki96} have already shown earlier how it is possible
within a 3D framework to equivalently reformulate the (unclosed)
turbulent balance equations of classical continuum mechanics into
a manifest form-invariant set of relations.}

Hence, fully in accord with Kretschmann's objection, we see that
no new physics is implied when reformulating the existing laws of
classical fluid motion into their generally form-invariant form.
This insight inevitably brings about the question of what the
necessity or even the advantage is when rewriting existing laws
into their general form-invariant representation? The answer
surely depends on what one intends to do. If for example the
equations are to be solved numerically or even analytically such
reformulations do not bring any advantages at all, but if for
example the equations are not closed and need to be modelled,
e.g., as the constitutive equations for materials or the equations
of turbulence, such form-invariant reformulations automatically
bring along consistent and structured modelling arguments in the
most natural way when they are based on invariant principles
\citep{Frewer09.2,Frewer09.3,Ariki15,Van15,Panicaud16}.

\subsection[The criticism on frame-indifference as a principle of nature]
{The criticism on frame-indifference as a principle of
nature\label{Sec4.2}}

The idea to pose frame-indifference as a principle of nature has
its formal roots in rational continuum mechanics as introduced by
\cite{Truesdell65}, in particular from the question how to
systematically model constitutive equations for materials in solid
and fluid mechanics.\footnote[3]{Historically, Hooke in 17th and
Poisson and Cauchy in the 19th century were the first to be
inspired by their experiments that the response of materials can
show frame-indifferent behavior. For more details on this account,
see e.g. Sec.$\,$2 in \cite{Frewer09.1} and the references
therein.} When put as a general principle to model the response of
a material, it states that only frame-indifferent (objective)
terms should enter the constitutive equation, independent of
whether an inertial or non-inertial system is considered. Hence in
the community of continuum mechanics this principle is also coined
as the principle of material frame-indifference~(MFI). Whether or
not MFI deserves the status of a principle in mechanics has been
subject of heated debates \citep{Frewer09.1}, and there still is a
need to clarify this issue
\citep{Dafalias11,Muschik12.1,Muschik12.2,Muschik12.3,Muschik13,Romano13,Liu14,Kirwan16,Yang16}.

For the discussion on MFI it is important to keep in mind that if
the constitutive equations for the considered material are
modelled explicitly as frame-indifferent, then this does not imply
that the dynamics of this material in space and time evolves
frame-indifferently too, since in general the dynamical equations
themselves are not free from frame-dependent (non-inertial)~terms.

Constitutive relations are of a different physical nature than the
field equations of motion into which they are inserted to obtain a
dynamically closed system. The former are modelling statements to
describe the response of a material to its intended environment,
while the latter constitute specific dynamical rules (driven by
conservation or balance laws) that determine the motion of this
material in this environment. For that difference, it can well be
that a material's response is frame-indifferent (objective), while
at the same time its dynamics evolves frame-dependently
(non-objectively). Hence, from the outset, it is not unreasonable
to assume MFI as a systematic modelling restriction for
constitutive relations.

The first controversy on MFI, however, was raised by
\cite{Mueller72}. By taking into account the microscopic
description of the material based on the kinetic theory of
Boltzmann, Müller argued that this theory does not support MFI,
according to which constitutive functions must be generally
frame-independent in all reference systems, including the
non-inertial systems. In particular, Müller concluded that MFI
cannot be regarded as an absolute principle. It rather has to be
seen only as an approximative principle
\citep{Mueller72,Mueller76}: Depending on the numerical ratio of
the microscopic time scale $t_\mu$ (defined proportional to the
mean free path length of the internal colliding constituents) of
the material and the macroscopic time scale $t_M$ of the
non-inertial frame environment, one can systematically determine
and decide whether MFI gives a good or a poor approximation for
the material within the environment considered. If there is a
clear-cut separation of time scales the decision is easy: a
sufficiently small ratio $t_\mu/t_M\ll 1$ certifies a very good
approximation, while a very large ratio $t_\mu/t_M\gg 1$ a poor
approximation. For solids and ordinary dense fluids the
MFI-principle is a valid principle to model their constitutive
relations (of course only as long as the macroscopic time scale
$t_M$ of the non-inertial frame environment does not approach a
relativistic regime). This, however, is no longer the case for
materials if the characteristic size of their microstructure gets
bigger as in suspensions or polymer solutions, or if the mean free
path length in the microprocesses gets longer as in gases and
rarefied gases. In all these cases MFI may not be used as a
guiding modelling principle~anymore.

However, with the subsequent development of extended
thermodynamics as a new systematic procedure to model constitutive
equations, new insight is provided into the principle of MFI.
According to Müller, the frame sensitivity of the stress and heat
flux discussed by him in \cite{Mueller72,Mueller76} can now
ultimately be explained on a pure macroscopic level without the
need to delve into a microscopic description of the problem
\citep{Mueller83}: It is only the natural frame dependence of the
basic balance equations for stress and heat flux which solely give
rise to the observed frame-dependence, not the constitutive theory
itself, which, according to him, still is frame-indifferent. In
other words, the microscopic analysis performed in
\cite{Mueller72,Mueller76}, particularly on the time scale ratio
$t_\mu/t_M$, only measures the effect and the relative strength of
frame-dependence in its entirety, but does not make any statements
from where this frame-dependence has its origin.

So what happened? On the one side Müller rejects MFI
\citep{Mueller72,Mueller76}, while on the other side he later
agrees on it \citep{Mueller83}. The answer depends on how a
constitutive equation or relation is defined. In classical or
ordinary thermodynamics, when based on the (unclosed) hierarchy of
moments of kinetic theory, the constitutive relations are
identified as fluxes in an iterative Maxwell scheme\footnote[2]{As
noted in \cite{Mueller72}, this naming goes back to
\cite{Ikenberry56} who invented this iterative scheme, which they
named after Maxwell, since Maxwell had sketched the beginning of
it: In an iterative manner this scheme determines the constitutive
relation between the fluxes and the densities (the field
variables) of the balance equations within some microscopic
interaction model (mostly Maxwellian molecules) for the underlying
probability density function obeying the Boltzmann
equation.\label{N2}} to close a fixed set of balance equations, as
was done in \cite{Mueller72,Mueller76}, while in extended
thermodynamics the constitutive relations are identified in a
first step as densities satisfying their own balance equations,
extending thus the fixed set of balance equations to higher order,
and then, in a second step, again as Maxwell iterates in order to
establish back a relationship to the constitutive relations of
ordinary thermodynamics by approximating the higher-order balance
equations near the equilibrium state of the fields. Hence,
according to Müller, the procedure of extended thermodynamics
allows to ``see this violation of material frame-indifference in a
new light" \citep[p.$\,$330]{Mueller83}. As defined and briefly
reviewed in Appendix~\ref{SecD}, classical thermodynamics operates
(after a second order Maxwell iteration) with constitutive
relations of the form
\begin{equation}
\vF=\vF^{(c)}(\vphi,\nabla\vphi,\nabla^2\vphi,\vGamma),\label{161102:1103}
\end{equation}
where $\vphi$ are the densities (mostly mass, momentum and
internal energy) and $\vF$ the fluxes (mostly stress and heat) of
the considered system. The parameters $\vGamma$ collectively
denote all possible frame-dependent quantities which arise as soon
as the considered system is transformed into a new one via a
coordinate transformation (see Appendix~\ref{SecD}). Hence, the
parameters $\vGamma$ appearing in the material law
\eqref{161102:1103} ultimately characterize the state of the
material if observed from a new frame of reference or if
transformed into a new frame of state. Extended thermodynamics,
however, operates with higher-order constitutive relations of the
form
\begin{equation}
\vJ=\vJ^{(e)}(\vphi,\vF),\qquad
\vXi=\vXi^{(e)}(\vphi,\vF),\label{161102:1122}
\end{equation}
where the higher-order or so-called extended fluxes $\vJ$ then
depend on the densities $\vphi$ and the fluxes $\vF$ of classical
thermodynamics, but at the price of gaining new unclosed
production terms $\vXi$ in the balance laws for $\vJ$. In contrast
now to classical thermodynamics, the modelling procedure in
extended thermodynamics only allows for local and
frame-indifferent constitutive relations as shown in
\eqref{161102:1122}; gradients and frame-dependent parameters do
not enter the higher-order constitutive relations
\eqref{161102:1122} as in \eqref{161102:1103} for classical
thermodynamics.

The standard argument for this difference in frame dependency for
the two types of constitutive relations is that the classical
non-local ones \eqref{161102:1103} are not true constitutive
equations but only approximations of balance laws, and hence do
not need to satisfy MFI, while the local ones~\eqref{161102:1122}
are true constitutive relations that have to obey MFI. In the
words of \cite{Ruggeri15}, p.$\,$360: ``Extended thermodynamics
seems to indicate in clear manner that nonlocal relations are not
constitutive equations but approximations of balance laws. The
true constitutive equations are in local form and they obey the
material frame difference". However, this argument of extended
thermodynamics, favored by \cite{Mueller98}, \cite{Liu14} and
\cite{Ruggeri15}, namely to push MFI in its {\it general}
statement as an axiom of nature for local constitutive equations,
is not convincing when considering the following three aspects:

\vspace{1em} \emph{(i)\label{(i)}} As explained and discussed in
detail by \cite{Muschik12.1,Muschik12.2,Muschik13}, any
constitutive law, does not matter if local or non-local, will pick
up frame dependent terms as soon as the balance equations are
formulated or transformed to non-inertial systems. That means for
any non-inertial system all constitutive relations in nature
inherently show explicit frame dependence through a collection of
parameters characterizing this particular non-inertial system,
which we denoted as~$\vGamma$ in \eqref{161102:1103} and which
\cite{Muschik12.1,Muschik12.2,Muschik13} calls the ``second entry"
in constitutive mappings to correctly describe the motion of the
material within any non-inertial system. In particular, this
``second entry" emerges independently of whether the material is
passively observed {\it from} a non-inertial system or whether the
material itself is actively transformed {\it into} a
non-inertial~system.

However, as also further correctly addressed by
\cite{Muschik12.1,Muschik12.2,Muschik13}, in order to ensure
physical consistency of this non-inertial frame dependency
$\vGamma$, all balance laws and all constitutive relations must be
formulated form-invariantly as true geometrical tensor relations,
in fact for all physically realizable frames, independent of
whether an inertial or non-inertial one is considered. Only this
will ensure that the motion of the material is described by the
same laws for each observer. In this respect, it is important to
note that the notions ``objective", ``tensor",
``observer-independence" and ``observer-invariance" are all used
as synonyms in \cite{Muschik12.1,Muschik12.2,Muschik13} to declare
the general notion of form-invariance, implying that {\it all}
observers will see the same laws of physics, i.e., where no
observer is distinguished, not even the inertial one.

\vspace{1em}\emph{(ii)} In order to satisfy MFI under Euclidean
(time-dependent Galilei) transformations, the local constitutive
relations \eqref{161102:1122} of extended thermodynamics are
standardly restricted to be isotropic functions of their variables
and not to depend on the velocity field
\citep[p.$\,$58]{Mueller98}. The latter restriction, however, is
too strong as it would unnecessarily exclude any velocity field
dependence {\it per se}. As frame-indifference restrictions under
Galilei transformations for inertial systems already show, it is
not necessary to exclude the velocity entirely, but that, instead,
it is fully sufficient to only consider velocity differences
relative to some fixed reference velocity of the system (see
Section \ref{Sec3}, in particular Eq.$\,$\eqref{161101:1018}).
When transforming to a non-inertial system, then it is this
velocity difference in the constitutive relations for example
which will give raise to frame-dependent terms as shown in the
last contribution of Eq.$\,$\eqref{161113:1849} and as explained
above in {\it (\hyperref[(i)]{i})}, independent of whether a
non-local or local constitutive relation is considered, since in
continuum mechanics the velocity field is defined as a local
(non-gradient) quantity and thus may enter both constitutive
relations, but, as said, as a velocity difference only. Hence, to
state that ``in extended thermodynamics with its local and
instantaneous constitutive functions material frame indifference
under [non-inertial] Euclidean transformations provides no more
restrictive conditions than material frame indifference under
[inertial] Galilean transformations does"
\citep[p.$\,$73]{Mueller98}, is not correct.

\vspace{1em}\emph{(iii)} In all of this discussion on MFI, one
always has to keep in mind that when regarding frame-indifference,
i.e., when regarding true objectivity and not the weaker
form-invariance, there only exists an equivalence principle for
inertial frames in nature, but not for non-inertial ones. In other
words, frame-indifference for physical laws can only be demanded
for inertial systems, a physical principle which applies without
exception. In this sense the inertial frame is distinguished from
all other frames: While the weaker principle of form-invariance
has to apply for all physically realizable frames, i.e. for
inertial as well as non-inertial ones, the stronger principle of
frame-indifference (true objectivity), however, may only be
applied to inertial frames and not to non-inertial ones. How
inertial frame indifference expresses itself or how it should be
enforced in physical models will be discussed in the next section.

It is still a mystery of nature why the inertial frame is
particularly distinguished from all other frames.\footnote[2]{In
reality the inertial frame is only an approximation, whereby the
error, of course, can be very small. For example, on Earth a true
inertial system cannot be established or generated due to its own
rotation, in planetary range outside Earth also not since our
Solar System rotates around the center of the Milky Way Galaxy,
and so~on.\linebreak Nevertheless, one cannot exclude that our
universe exhibits some local inertial regions in deep space that
are close to perfection.} For example, the distinction between an
inertial and an accelerated system is not yet fully understood.
The question remains of relative to what an inertial system is not
accelerated. Only to say that it moves with a constant velocity
relative to a sky of fixed stars is certainly not precise enough.
This discussion was first introduced by \cite{Mach83} and is still
an open discussion today
\citep{Friedman83,Barbour95,Barbour04,Penrose05,Pfister15}. Mach's
ideas cumulated to {\it Mach's Principle}, a name given by
Einstein which served him as a guiding criterion to develop his
general theory of relativity. Today we understand Mach's principle
as a hypothesis that all mass of the universe determines the
structure and behavior of an inertial system. The question,
however, in how far this principle is incorporated or reflected by
Einstein's theory is still open.

\newpage
\subsection{Formulating a universal axiom for invariant modelling\label{Sec4.3}}

Accounting for all three aspects mentioned above, we can state for
general continuum mechanics the following \textbf{reduced axiom of
Material Frame-Indifference (r-MFI):}\footnote[2]{The add-on
`reduced' refers here to the fact that MFI, regarding its
ingredient `frame-indifference', is only considered as a general
principle for inertial Galilean transformations, and not for the
non-inertial Euclidean transformations as standardly and
incorrectly assumed in the community of continuum mechanics. Of
course, assuming MFI for Euclidean transformations can be a good
approximation for some practical problems in engineering
mechanics, since mostly in these cases the influence of the
material's motion (either induced through an active or passive
coordinate transformation) on the constitutive equation can be
neglected \citep{Muschik12.1}.}
\vspace{-0.25cm}\begin{enumerate}[label=\arabic*.,leftmargin=2em]
\setlength\itemsep{-0.25em}
\item[(I)] {\it Regarding form-invariance (covariance)}, all
balance equations and all constitutive relations have to be
formulated as tensor relations for all physically realizable
frames in order to satisfy the general principle of
form-invariance so that no observer is distinguished, not even the
inertial one. The most natural framework to achieve this is to
formulate all physical laws in a 4D space-time setting. How to
stay in the realm of classical Newtonian physics, two approaches
exist: Either by formulating directly a 4D Newtonian manifold
endowed with a time-like and a space-like metric
\citep{Havas64,Matolcsi07,Frewer09.1,Frewer09.2,Frewer09.3,Van15},
or indirectly by using Einstein's space-time theory in the
classical limit of small velocities with respect to the speed of
light \citep{Rouhaud13,Panicaud16}.
\item[(II)] {\it Regarding frame-indifference (objectivity)}, the
inertial frame is distinguished from all other frames in physics
since only for these a general equivalence principle exists: All
inertial frames are indistinguishable. The connecting coordinate
transformations are the Galilei transformations in classical and
the Lorentz transformations in relativistic continuum mechanics.
For balance equations and constitutive relations this equivalence
principle expresses itself differently:\newline $(a)$ Balance
equations are to be formulated frame-indifferently under Galilei
or Lorentz transformations only for closed systems, i.e., if no
external body forces act on the considered thermodynamic system,
since such forces may break the space-time symmetries necessary
for Galilean or Lorentzian invariance. Practically this is
achieved in three steps, first by deactivating all external forces
in the balance laws by putting them formally to zero, then, in the
second step, restricting these resulting balance laws to Galilean
frame-indifference, and then, in the last step, to re-activate all
external forces back to their original form again.\newline $(b)$
The constitutive relations will automatically be restricted to
Galilean frame-indifference as soon as the above process $(a)$ is
initiated. As a consequence, the constitutive equations will not
depend on any absolute locations in space and time, nor on any
absolute directions in space nor on any absolute velocities. If
modelled as such, only relative distances, times differences,
directional differences and relative velocities enter the
constitutive relations (see Section~\ref{Sec3}), independent of
whether the modelling concept of classical or extended
thermodynamics is employed. Of course, the latter concept is
superior to the former one in that it not only adopts more
independent variables by incorporating non-equilibrium variables
such as viscous stress and heat flux into the theory, but also by
modelling the constitutive relations locally and instantaneously
without gradients, having the twofold effect then that on the one
side the theory is governed by hyperbolic field equations allowing
only for\linebreak finite speeds of propagations, and, on the
other side, having simple access to the principle of
frame-indifference for closed inertial systems.
\end{enumerate}
\vspace{-1em} \PRLsep\vspace{-1em}\newline
Hence a thermodynamical system which is formulated according to
the above r-MFI principle will involve constitutive equations that
contain information about the motion of the material
\citep{Muschik12.2,Muschik13}, irrespective of whether this motion
is induced by observing the material passively from a new frame of
reference or by actively transforming the material into a new
frame of state. However, this frame-dependence or
frame-sensitivity of the material will only emerge when changing
to a non-inertial system, since for all inertial systems the
material behaves frame-\pagebreak[4] indifferently by
construction. In other words, all constitutive equations,
irrespective of whether local or non-local, will pick up
frame-dependent terms as soon as one changes to a non-inertial
system, reflecting essentially the material's sensitivity upon
acceleration. In \cite{Muschik12.2} this acceleration-sensitivity
of materials is regulated by the MMD-axiom, the axiom of `Material
Motion Dependence'.

To close this section, it is worthwhile to mention the note of
\cite{Muschik12.1,Muschik12.2}, that the existence of
acceleration-sensitive materials is well known in physics. A
famous example is the Barnett-effect \citep{Barnett15}, where upon
rotation a magnetization of an uncharged body is
induced;\footnote[2]{Obviously the Barnett-effect can only be
observed for magnetizable bodies, e.g. made of iron or nickel.}
irrespective of whether the body is passively observed from a
rotating frame or if actively put into rotation, the magnetization
is always proportional to the angular velocity and thus parallel
to the angular velocity of the body \citep{Matsuo15}. Hence, it
may well be that the Barnett-effect gives a relevant contribution
towards the Earth's magnetic field due to being in a state of
constant rotation.

\subsection{On form-invariance and frame-indifference in turbulence
modelling\label{Sec4.4}}

Using the concepts of form-invariance and frame-indifference in
turbulence modelling has its motivation from MFI when interpreting
the unclosed and modelled turbulent flow quantities as
constitutive relations which are necessary to close the underlying
dynamical equations of motion. Despite this similarity, it is
clear that physically one has to distinguish between modelling a
turbulent flow and modelling constitutive equations of a
continuous material: The closure problem in turbulence modelling
is a pure dynamical problem which originates from the lack of
knowledge in quantifying the complex flow behavior of a fluid
material, and not from the lack of knowledge concerning the
structure and behavior of the material itself. Yet, there is a
common link between these two entities, as correctly stated by
\cite{Dafalias11}: Both must be valid for {\it any} motion and in
regard to {\it any} frame, if they are to be something more than
just a convenient fitting for the solution of particular problems
associated with particular motions and particular frames. It is
exactly this quest of validity for ``any motion and any frame"
that makes it possible to impose objectivity requirements and, as
a result, draw conclusions for the appropriate analytical
representation of either constitutive or dynamical relations.

However, for the same reasons as stated before in the previous
section, it is clear that not MFI, but only r-MFI can apply to
turbulence modelling. But for turbulence modelling this limitation
is a stringent condition, in contrast to material modelling where
MFI can still be used as a reasonable approximation if the
influence of the material's motion on its constitutive equation
can be neglected \citep{Muschik12.1}. An indicator for frame
sensitivity in general is the time-scale ratio $t_\mu/t_M$
introduced in Section \ref{Sec4.2}, to compare the microscopic
time scale of the material's constituents with the macroscopic
time scale of its frame environment. Although this ratio measures
frame sensitivity only as a whole, not distinguishing between the
frame-dependence originating from the balance laws or from the
constitutive relations, it nevertheless is a global indicator of
whether we face a weak ($t_\mu/t_M\ll 1$) or a strong
($t_\mu/t_M\gg 1$) dependence of the material on the frame.

Now, when applying this indicator to turbulent motion of a fluid,
\cite{Lumley70,Lumley83} correctly pointed out that a general
principle of frame-indifference, as stated by MFI, cannot be
utilized for turbulence modelling, not even in an approximative
sense as it can be done in rational continuum mechanics when
modelling the response of a material. The reason is simply the
huge spectrum of (microscopic) time-scales within a turbulent flow
where there always exists a band of scales which are in the
(macroscopic) time-scale range of the underlying non-inertial
frame of reference. In other words, in a turbulent flow there is
no clear-cut separation of microscopic and macroscopic
time-scales, as it would be the case, for example, in material
modelling when considering a {\it dense} material within a certain
{\it slow} changing non-inertial frame environment, where
obviously, due to $t_\mu/t_M\ll 1$, frame-indifference as stated
by MFI can be used as a reasonable guiding principle to model such
materials in such environments.

Hence, since MFI may not be used as a guiding principle to model
turbulent flows, not even in an approximative sense, makes the
role of r-MFI in turbulent flows even more explicit. But as
defined and formulated in the previous section, is r-MFI really a
sufficient principle to account for all invariant aspects of a
turbulent flow? Particularly in view of the fact that in contrast
to material modelling, turbulence is based on an underlying
deterministic description where the direct equations of motion are
known exactly --- in classical fluid mechanics, these are the
Navier-Stokes equations, where, in the following, we will only
consider the incompressible approximation.

It is clear that r-MFI forms the basis as a principle, providing a
rule of procedure that strictly applies to any modelling ansatz in
physics, without exception. But, particularly for turbulence
modelling, where there is a direct link down to the underlying and
exactly known microscopic (fine-grained) equations of motion, the
central question will be, if r-MFI can be extended with regard to
its second part (II), the frame-indifference (objectivity)
principle, to include more symmetries than just the Galilean (or
Lorentzian) symmetry? The answer is yes and no, which will be
discussed next.

\vspace{-0.1cm}\section{Supplementing r-MFI to model
incompressible turbulent flow\label{Sec5}}

Although the study of \cite{Speziale98} shows the serious drawback
that it can be misunderstood to due a systematic confusion of the
concepts `form-invariance' and
`frame-indifference',\footnote[2]{For example, Speziale leaves the
reader with the following confusing statement: ``The fact that the
Reynolds stress tensor is a frame-indifferent tensor does not
establish the validity of Material Frame-Indifference"
\cite[p.$\,$498]{Speziale98}; a statement which unfortunately
generated misguidance as for example in study of
\cite{Dafalias11}. The problem is that Speziale understands
`form-invariance' as `frame-indifference', by showing that under
arbitrary 3D Euclidean (non-inertial) transformations the
Reynolds-stress transforms as a tensor of rank 2 , namely
as\linebreak $\vtau^*=\vQ\vtau\vQ^T$
\cite[Eqs.$\,$(70)-(74)]{Speziale98}. However, this just shows
form-invariance (covariance) of the Reynolds-stress $\vtau$, and
not its frame-indifference (objectivity). Hence Speziale's
confusing statement has to be correctly read as: ``The fact that
the Reynolds-stress is a tensor, i.e. form-invariant, does not
establish the validity of Material Frame-Indifference"; a
statement which now turns out to be a trivial statement from the
viewpoint of the notions defined herein in this study (see e.g.
second footnote on p.$\,$\pageref{N4}).} it still marks the key
study in how turbulence should be modelled when based on invariant
principles. Basically this study addresses two points: Firstly,
MFI under 3D Euclidean coordinate transformations, as generally
stated in rational continuum mechanics, does not apply to
turbulence, and secondly, in order to model turbulence according
to the principles of invariance correctly, it is necessary to also
include the fine-grained (microscopic or fluctuating) description
of turbulence, saying that it is not expedient to focus solely on
the coarse-grained (macroscopic or mean) description of
turbulence.

Speziale is thus in accordance with the criticism of
\cite{Lumley70,Lumley83} as it was discussed in the previous
section,\footnote[3]{Speziale's work is to be separated into two
classes, those studies concerning material modelling where he
favors MFI as a valid modelling principle in its full 3D original
formulation \citep{Speziale84,Speziale87,Speziale98}, and those
studies concerning turbulence where he denies it
\citep{Speziale89,Speziale91.2,Speziale98}. It is not true that
Speziale favored full MFI in turbulence modelling before writing
his last review \citep{Speziale98} and comment \citep{Spalart99},
except for his very first study on this issue \citep{Speziale79},
where took a contrary position. He always stated that MFI in
turbulence cannot be valid in~3D, but only in 2D under certain
restrictions. However, when excluding 3D rotation, he constantly
supplemented his statement by the argument that, nevertheless, in
the specific case of linear accelerations a 3D-MFI for turbulence
can be formulated according to an extended set of Galilei
transformations.} yet he goes beyond this criticism in asking, if
not MFI, what then are the correct invariance principles for
turbulence? The solution to this problem is to look at the
symmetry properties of the transport equations for the fluctuating
variables, since the unclosed terms are just an averaged
multiplicative combination of these, or in his words: ``Since
closure relations are built up from solutions of the fluctuation
dynamics, we must look at its invariance groups"
\cite[p.$\,$498]{Speziale98}.\pagebreak[4]

However, in this step caution has to be exercised, because in
order to avoid any wrong interpretations and conclusions from the
outset, it proves to be useful {\it not} to analyze the pure
fluctuating but rather the full instantaneous equations for
symmetries, which include both the mean and fluctuating dynamics
of the flow. The simple reason is that if the symmetry analysis is
not carried out careful enough by providing and revealing all
information originating from the pure fluctuating dynamics, the
governing equations will turn into an unclosed system due to the
appearance of an infinite hierarchy of statistical equations
triggered by the mean velocity field. A condition which is
obstructive to any symmetry analysis, because, as explicitly shown
e.g.~in \cite{Frewer15.0,Frewer15.0x,Frewer16.3,Frewer16.4}, to
determine the symmetries for such an unclosed and infinite
equational system is basically ill-defined: If the equations are
unclosed, so are their symmetries.\footnote[2]{Note that to use
the word `symmetry' in the context of unclosed systems is
mathematically incorrect: Any invariant transformation admitted by
an unclosed system of equations can only be identified as an
equivalence transformation and not as a true symmetry
transformation; see e.g. \cite{Frewer14.1} and the
references~therein.} Performing a symmetry analysis on such
unclosed and infinite systems has the negative consequence that
one is not only faced with complete arbitrariness in generating
symmetries, but that also unphysical symmetries in this process
can be generated which violate the classical principle of cause
and effect
\citep{Frewer14.1,Frewer15.1,Frewer16.1,Frewer16.2,Frewer16}.\footnote[3]{To
the peer-reviewed publications
\citep{Frewer14.1,Frewer15.1,Frewer16.2,Frewer16}, please also see
the reactions \citep{Frewer14.0,Frewer.X2,Frewer.X3,Frewer.X4},
respectively.}

Now, in order to extend the second part (II) of the axiom r-MFI,
one has to determine all invertible coordinate transformations
(within Newtonian physics)
\begin{equation}
\tilde{t}=t+c_0,\qquad
\tilde{\vx}=\tilde{\vx}(\vx,t),\label{161106:2139}
\end{equation}
that leave the incompressible Navier-Stokes equations
\begin{equation}
\nabla\cdot \vu =0,\qquad \partial_t \vu
+\vu\cdot\nabla\vu=-\nabla p+\nu\Delta\vu,\label{161106:2141}
\end{equation}
frame-indifferent. Important to note here is that for pure spatial
(time-independent) transformations $\tilde{\vx}=\tilde{\vx}(\vx)$,
equations \eqref{161106:2141} are 3D-form-invariant if all spatial
operators are identified as 3D covariant
derivatives.\footnote[4]{See Appendix~\ref{SecB} for a general
definition of the covariant derivative \eqref{161121:1243}, where
the Greek 4D indices have to be replaced by the Latin 3D indices
in order to obtain a 3D spatial covariant derivative. For example,
in \eqref{161106:2141} the 3D-form-invariant continuity equation
then has the explicit componential form
$$0=\nabla\cdot\vu=\nabla_i u^i=\partial_i u^i+\Gamma^i_{ik}u^k=
\frac{1}{\sqrt{\lvert g\rvert}}\frac{\partial}{\partial
x^k}\left(\sqrt{\lvert g\rvert}\, u^k\right),$$ where
$\Gamma^{i}_{jk}$ is the affine connection and $g=\det (g_{ij})$
the determinant of the metric tensor $g_{ij}$ of the considered 3D
spatial manifold.} However, a consistent 3D tensor relation of
\eqref{161106:2141} is only warranted when the pressure term
$\nabla p$ is interpreted as a contravariant gradient with
components\linebreak $(\nabla p)^i=\nabla^i p=g^{ij}\nabla_j p$,
where $g^{ij}$ are the contravariant coefficients of the metric
tensor with the standard defining property
$g^{ik}g_{kj}=\delta^i_j$, and, of course, the convective term
as~$\vu\cdot\nabla\vu\mathrel{\widehat{=}}
(\vu^T\cdot\nabla^T)\vu$.

Not form-invariance, but frame-indifference under
\eqref{161106:2139} for \eqref{161106:2141} is now given if this
transformation results to
\begin{equation}
\tilde{\nabla}\cdot \tilde{\vu} =0,\qquad \partial_{\tilde{t}}
\tilde{\vu}
+\tilde{\vu}\cdot\tilde{\nabla}\tilde{\vu}=-\tilde{\nabla}
\tilde{p}+\nu\tilde{\Delta}\tilde{\vu},\label{161106:2147}
\end{equation}
where (i) the metric tensor is kept invariant
$\tilde{g}_{ij}=g_{ij}$, and (ii) where the transformation rules
for the fields $\vu\sim \dot{\vx}=d\vx/dt$ and $p$ are induced by
\eqref{161106:2139} as
\begin{equation}
\tilde{\vu}=\frac{\partial\tilde{\vx}}{\partial\vx}\cdot
\vu+\frac{\partial\tilde{\vx}}{\partial t},\qquad
\tilde{p}=p,\label{161106:2246}
\end{equation}
being the only physically consistent transformation rule for the
velocity field $\vu=\vu(\vx,t)$ and the pressure field
$p=p(\vx,t)$, when changing the frame of reference according to
\eqref{161106:2139}. Note that if these two fields are known in
the initial frame, then \eqref{161106:2246} represents the
explicit construction rule for these fields in the transformed
frame
\begin{equation}
\tilde{\vu}(\tilde{\vx},\tilde{t})\equiv\frac{\partial\tilde{\vx}}{\partial\vx}
\bigg|_{\vx=\vx(\tilde{\vx},\tilde{t});\: t=\tilde{t}-c_0}\cdot
\vu\big(\vx(\tilde{\vx},\tilde{t}),\tilde{t}-c_0\big)+\frac{\partial\tilde{\vx}}{\partial
t} \bigg|_{\vx=\vx(\tilde{\vx},\tilde{t});\: t=\tilde{t}-c_0}
\end{equation}
and
\begin{equation}
\tilde{p}(\tilde{\vx},\tilde{t})\equiv
p\big(\vx(\tilde{\vx},\tilde{t}),\tilde{t}-c_0\big).
\end{equation}
The reason why we only look at pure coordinate transformations
\eqref{161106:2139} when trying to extend\linebreak axiom r-MFI,
is that it's a principle that by construction only relates to
coordinate transformations, in particular since r-MFI is based on
the concept of tensors, which is a well-defined concept only for
coordinate transformations concerning the change of a considered
reference~frame.

For simplicity, we will formulate the symmetry results for the
Navier-Stokes equations \eqref{161106:2141} only in 3D Cartesian
(rectilinear) coordinates, i.e., in the following only for
$g_{ij}=\delta_{ij}$. It is clear that for all other spatial
(curvilinear) coordinate transformations the symmetry property of
the Navier-Stokes equations remains unchanged, since a pure
spatial coordinate transformation in 3D only results into a pure
relabelling action of the coordinates, not changing the physics of
the Navier-Stokes equations \eqref{161106:2141}, due to being
formulated as a 3D-form-invariant tensor relation. Hence, each
symmetry transformation that will be listed below will represent
an independent symmetry of the Navier-Stokes system modulo (up to)
all its relabelling symmetries.

Performing a systematic Lie-group symmetry analysis on the
incompressible Navier-Stokes equations \eqref{161106:2141}, as
first done by \cite{Danilov67}, \cite{Bytev72} and
\cite{Pukhnachev72}, reveals that the only set of continuous
coordinate transformations admitted as a symmetry are the Galilei
transformations
\begin{equation}
\mathsf{G}:\quad \tilde{t}=t+c_0,\quad \tilde{\vx}=\vA \vx +\vc_1
t +\vc_2\quad\underset{\eqref{161106:2246}}{\Rightarrow}\quad
\tilde{\vu}=\vA \vu+\vc_1,\quad\tilde{p}=p, \label{161107:1304}
\end{equation}
where $\vA$ is a constant rotation matrix and $c_0$, $\vc_1$,
$\vc_2$ arbitrary constants. The two remaining continuous
symmetries of the 3D Navier-Stokes equations
\begin{equation}
\left.
\begin{aligned}
&\mathsf{S}_{\mathsf{1}}:\quad \tilde{t}=e^{2\varepsilon}t,\;\;\;
\tilde{\vx}=e^{\varepsilon}\vx,\;\;\;
\tilde{\vu}=e^{-\varepsilon}\vu,\;\;\;\tilde{p}=e^{-2\varepsilon}p,\\[0.5em]
&\mathsf{S}_{\mathsf{2}}:\quad \tilde{t}=t,\;\;\;
\tilde{\vx}=\vx+\vf(t),\;\;\;
\tilde{\vu}=\vu+\dot{\vf}(t),\;\;\;\tilde{p}=p-\vx^T\cdot
\ddot{\vf}(t)+g(t),\;\;\text{for}\;\; \ddot{\vf}(t)\neq
\boldsymbol{0},
\end{aligned}
~~~\right\}\label{161107:2033}
\end{equation}
are {\it no} coordinate transformations, since the fields do not
transform as \eqref{161106:2246}. In particular for the so-called
`extended Galilei transformation' $\mathsf{S}_{\mathsf{2}}$, the
pressure is not transforming as a scalar function, i.e., not
invariantly and thus generating a frame-dependence in the solution
of the equations: If the pressure is (locally) zero in one frame
$p=0$, then it is not zero in another frame $\tilde{p}\neq 0$, and
vice versa, thus distinguishing (locally at a certain point) that
frame where the pressure is zero from all other frames where it is
not zero.\footnote[2]{Note that, in contrast to the pressure field
$p$, the velocity field $\vu$ in $\mathsf{S}_{\mathsf{2}}$
transforms form-invariantly as \eqref{161106:2246}; its
form-invariant tensor structure, however, can only be explicitly
seen in a 4D formulation
\citep{Frewer09.1,Frewer09.2,Frewer09.3}.} It is thus misleading
to call $\mathsf{S}_{\mathsf{2}}$ an extension of the Galilei
transformation $\mathsf{G}$ \eqref{161107:1304}, as standardly
done in the community of fluid mechanics (see e.g.
\cite{Speziale98}), because $\mathsf{S}_{\mathsf{2}}$ is
physically of a different nature than $\mathsf{G}$, being a
coordinate transformation while $\mathsf{S}_{\mathsf{2}}$ not.

It is exactly in this sense that r-MFI cannot be extended to
include more symmetries than the Galilei transformation
($\mathsf{G}$), since all other continuous symmetries of the
Navier-Stokes equations ($\mathsf{S}_{\mathsf{1}}$,
$\mathsf{S}_{\mathsf{2}}$) are no coordinate transformations for
which r-MFI applies. But the symmetries \eqref{161107:2033} cannot
be ignored, since they reflect a characteristic physical behavior
of the Navier-Stokes equations: $\mathsf{S}_{\mathsf{1}}$ ensures
the correct scaling of all physical dimensions involved in the
flow, while  $\mathsf{S}_{\mathsf{2}}$ is a re-gauging of the
pressure field when changing to a linearly accelerated frame of
reference. Hence, when modelling turbulent flow according to the
incompressible Navier-Stokes equations, the axiom r-MFI needs to
be supplemented by a second, independent axiom, to be called:

\vspace{1em}\noindent\textbf{Turbulent Frame-Indifference (TFI):}
All turbulence models have to be consistent with the\linebreak
invariant properties of the Navier-Stokes equations (modulo
relabelling). In particular, besides the principle of r-MFI, all
turbulence models have to respect the two continuous
symmetries~\eqref{161107:2033}, along with the two further
discrete symmetries$\,$\footnote[2]{Mathematically correct, the
invariance $\mathsf{S}_{\mathsf{4}}$ in \eqref{161107:2141} is not
a symmetry, but only an equivalence transformation (see e.g.
Sec.$\,$2 in \cite{Frewer14.1} and the references therein).
Although this transformation is not physically realizable, since
it is impossible to construct a fluid having negative molecular
viscosity, it nevertheless represents a mathematical property
which is characteristic for the Navier-Stokes equations. In
particular this equivalence transformation will serve as guideline
when developing, e.g., a molecular viscosity expansion within a
certain turbulence model \citep{Frewer09.3}. In this regard, it
should be noted that the discrete equivalence
$\mathsf{S}_{\mathsf{4}}$ is not the only equivalence
transformation which the Navier-Stokes equations admit when trying
to change the value of the molecular viscosity; because from the
viewpoint of a continuous transformation, an infinite Lie-algebra
of such equivalence transformations exist \citep{Unal94,Unal95}.}
of the 3D Navier-Stokes equations \eqref{161106:2141}
\begin{equation}
\left.
\begin{aligned}
&\mathsf{S}_{\mathsf{3}}:\quad \tilde{t}=t,\;\;\;
\tilde{x}^i=-x^i,\;\;\;
\tilde{x}^j=x^j,\;\;\;\tilde{u}^i=-u^i,\;\;\;
\tilde{u}^j=u^j,\;\;\;
\tilde{p}=p,\;\;\;\text{with}\;\; j\neq i,\\[0.5em]
&\mathsf{S}_{\mathsf{4}}:\quad \tilde{t}=-t,\;\;\;
\tilde{\vx}=\vx,\;\;\;
\tilde{\vu}=-\vu,\;\;\;\tilde{p}=p,\;\;\;\tilde{\nu}=-\nu.
\end{aligned}
~~~\right\}\label{161107:2141}
\end{equation}
Additional symmetries may also be utilized for turbulence
modelling, but only in an approximative sense. For example, if in
the turbulent flow domain there is a (localized) region where the
influence of viscosity on the flow is negligibly small, then,
independent from $\mathsf{S}_{\mathsf{1}}$, a second, but
approximative scaling symmetry can be considered as a modelling
restriction, namely the additional 3D scaling symmetry admitted
exactly only by the inviscid $(\nu=0)$ Euler equations
\begin{equation}
\mathsf{S}^{\mathsf{(approx.)}}_{\mathsf{5}}:\quad\tilde{t}=t,\;\;\;
\tilde{\vx}=e^{a}\vx,\;\;\;\tilde{\vu}=e^{a}\vu,\;\;\;
\tilde{p}=e^{2a}p,
\end{equation}
or, if some acting body force in the system drives the 3D
turbulent flow into a 2D state, as for example the Coriolis force
in a rapidly rotating turbulent flow in the long time limit
\citep{Gallet15,Machicoane16}, then an approximative uniform
rotation symmetry can be considered, admitted exactly only by the
2D Navier-Stokes equations
\begin{align}
\mathsf{S}^{\mathsf{(approx.)}}_{\mathsf{6}}:&\quad\tilde{t}=t,\;\;\;
\tilde{x}^i=Q^i_{\,\, j}x^j,\;\;\;\tilde{u}^i=Q^i_{\,\,
j}u^j+\dot{Q}^i_{\,\,
j}x^j,\;\;\text{for}\;\; i,j=1,2,\nonumber\\[0.5em]
&\quad\text{and}\;\;
\tilde{p}=p+{\textstyle\frac{1}{2}}\omega_z^2\delta_{ij}x^ix^j+
2\omega_z\psi,\;\;\text{with}\;\;
\psi=-\int_C\left(u^2dx^1-u^1dx^2\right),\label{161114:2037}
\end{align}
where $Q^i_{\,\, j}$ represents a two-dimensional uniform rotation
matrix with $\omega_z$ being the constant angular velocity and
$\psi$ the corresponding stream function defined as a planar curve
integral over the two-componential velocity field $u^i$. This
exact symmetry of the 2D Navier-Stokes equations was first derived
independently by \cite{Pukhnachev60,Cantwell78,Speziale81}, and it
was Speziale who misleadingly coined this invariance as `2D-MFI',
which is misleading in so far as the symmetry
$\mathsf{S}_{\mathsf{6}}^{\mathsf{(approx.)}}$ does not form a
coordinate but only a re-gauging transformation in the same way as
discussed before for the non-coordinate transformation
$\mathsf{S}_{\mathsf{2}}$ \eqref{161107:2033}.
\newline\vspace{0em} \PRLsep\vspace{0em}\newline
To see how the two invariance principles r-MFI and TFI act as
modelling restrictions for turbulence, let us for simplicity first
consider the option to model the unclosed Reynolds stresses
algebraically, i.e., directly as a constitutive equation in the
analogous way as it is done in classical or ordinary
thermodynamics (see Appendix~\ref{SecD}), by relating the Reynolds
stresses to the fields and their gradients of the underlying
balance equations, here to the mean incompressible continuity and
momentum equations of \eqref{161106:2141}, which again, for
reasons of simplicity and clarity, will only be demonstrated
explicitly for Cartesian (rectilinear)
coordinates$\,$\footnote[3]{In the following, only the concept of
`frame-indifference' is considered, i.e., only the concept as how
to incorporate the symmetries of the Navier-Stokes equations into
the turbulence modelling process, and not so much the concept of
`form-invariance', which has been done and discussed elsewhere;
see e.g. \cite{Sadiki96,Frewer09.1,Frewer09.2,Frewer09.3,Ariki15},
where the form-invariant (covariant) Reynolds-averaged
Navier-Stokes are formulated and discussed both in 3D and 4D.}
\begin{equation}
\partial_i\L u^i\R=0,\qquad\partial_t \L u^i\R+\L u^j\R\partial_j \L
u^i\R=-\delta^{ij}\partial_j \L
p\R+\nu\delta^{jk}\partial^2_{jk}\L
u^i\R-\partial_j\tau^{ij},\label{161111:1256}
\end{equation}
where
\begin{equation}
\tau^{ij}=\L u^{\prime\hspace{0.25mm} i}u^{\prime\hspace{0.25mm}
j}\R = \lim_{N\to\infty}\frac{1}{N}\sum_{n=1}^N
u^{\prime\hspace{0.25mm} i}_{(n)}u^{\prime\hspace{0.25mm}
j}_{(n)},\label{161113:0838}
\end{equation}
are the Reynolds stresses defined as an ensemble average over $n$
flow realizations of the velocity product
$u^{\prime\hspace{0.25mm} i}u^{\prime\hspace{0.25mm} j}$ for the
fluctuating field $u^{\prime\hspace{0.25mm} i}=u^i-\L u^i\R$. Each
realization of the fluctuating velocity field satisfies the same
continuity and momentum equations obtained by subtracting the mean
Navier-Stokes equations \eqref{161111:1256} from the Reynolds
decomposed instantaneous ones \eqref{161106:2141}:
\begin{equation}
\left.
\begin{aligned}
\partial_i u^{\prime\hspace{0.25mm} i}_{(n)}=0,\hspace{5.25cm}\\[0.5em]
\partial_t u^{\prime\hspace{0.25mm} i}_{(n)}+ u^{\prime\hspace{0.25mm} j}_{(n)}
\partial_j  u^{\prime\hspace{0.25mm} i}_{(n)}+ \L u^j\R
\partial_j u^{\prime\hspace{0.25mm} i}_{(n)}+u^{\prime\hspace{0.25mm} j}_{(n)}
\partial_j \L u^i\R=-\delta^{ij}\partial_j
p^\prime_{(n)}+\nu\delta^{jk}\partial^2_{jk}u^{\prime\hspace{0.25mm}
i}_{(n)}+\partial_j\tau^{ij}.\label{161114:2034}
\end{aligned}
~~~ \right\}
\end{equation}
Formally the system \eqref{161111:1256}-\eqref{161114:2034}
constitutes a closed system of equations$\,$\footnote[2]{The fact
that system \eqref{161111:1256}-\eqref{161114:2034} can be
considered as formally closed, is due to taking along the
definition of the ensemble average \eqref{161113:0838}, which
forms the defining relation between the fine-grained (microscopic)
and the coarse-grained (macroscopic) description of turbulence. If
this information \eqref{161113:0838} is not provided to an
analysis, e.g., on symmetries, the mean equations
\eqref{161111:1256} obviously turn into an unclosed set of
equations, for which, as was already discussed before, a true
symmetry analysis is ill-defined due to the effect of having to
deal with an infinite hierarchy of statistical moment equations
triggered by the lowest order equation \eqref{161111:1256}.} which
admits the same set of symmetry transformations
\eqref{161107:1304}-\eqref{161114:2037} as the instantaneous
Navier-Stokes equations, however, now in the specific Reynolds
decomposed form$\,$\footnote[3]{Indeed, the transformations
\eqref{1611142101}-\eqref{161114:2111} are admitted as symmetries,
and \eqref{161114:2112}-\eqref{161114:2103} as approximative
symmetries of the combined system
\eqref{161111:1256}-\eqref{161114:2034}. A systematic Lie-group
symmetry analysis in fact even shows that the instantaneous
Navier-Stokes symmetries can only be decomposed into its mean and
fluctuating part as given in
\eqref{1611142101}-\eqref{161114:2103}. Any other decomposition
would not be admitted as a symmetry of this system
\eqref{161111:1256}-\eqref{161114:2034}.}
\begin{align}
\mathsf{G}:&\quad \tilde{t}=t+c_0,\quad \tilde{x}^i=A^i_{\,\,j}
x^j+c^i_1 t +c^i_2,\quad
\L\tilde{u}^i\R=A^i_{\,\, j}\L u^j\R+c^i_1,\quad\L\tilde{p}\R=\L p\R,\nonumber\\
&\quad  \tilde{u}^{\prime\hspace{0.25mm} i}_{(n)}=A^i_{\,\,j}
u^{\prime\hspace{0.25mm} j}_{(n)},\quad
\tilde{p}^\prime_{(n)}=p^\prime_{(n)},\label{1611142101}\\[0.5em]
\mathsf{S}_{\mathsf{1}}:&\quad \tilde{t}=e^{2\varepsilon}t,\quad
\tilde{x}^i=e^{\varepsilon}x^i,\quad \L
\tilde{u}^i\R=e^{-\varepsilon}\L u^i\R,\quad
\L\tilde{p}\R=e^{-2\varepsilon}\L p\R,\nonumber\\
&\quad \tilde{u}^{\prime\hspace{0.25mm}
i}_{(n)}=e^{-\varepsilon}u^{\prime\hspace{0.25mm} i}_{(n)},\quad
\tilde{p}^\prime_{(n)}=e^{-2\varepsilon}p^\prime_{(n)},\label{161114:2119}\\[0.5em]
\mathsf{S}_{\mathsf{2}}:&\quad \tilde{t}=t,\quad
\tilde{x}^i=x^i+f^i(t),\quad \L\tilde{u}^i\R=\L
u^i\R+\dot{f}^i(t),\quad \L\tilde{p}\R=\L p\R-\delta_{ij}x^i
\ddot{f}^j(t)+\L g(t)\R,\qquad\quad\nonumber\\
&\quad \tilde{u}^{\prime\hspace{0.25mm}
i}_{(n)}=u^{\prime\hspace{0.25mm} i}_{(n)},\quad
\tilde{p}^\prime_{(n)}=p^\prime_{(n)}+g^\prime(t),\quad\text{for}\;\;
\ddot{f}^i(t)\neq 0,\\[0.5em]
\mathsf{S}_{\mathsf{3}}:&\quad \tilde{t}=t,\quad
\tilde{x}^i=-x^i,\quad \tilde{x}^j=x^j,\quad\L\tilde{u}^i\R=-\L
u^i\R,\quad \L\tilde{u}^j\R=\L u^j\R,\quad \L\tilde{p}\R=\L
p\R,\nonumber\\
&\quad \tilde{u}^{\prime\hspace{0.25mm}
i}_{(n)}=-u^{\prime\hspace{0.25mm} i}_{(n)},\quad
\tilde{u}^{\prime\hspace{0.25mm} j}_{(n)}=u^{\prime\hspace{0.25mm}
j}_{(n)},\quad
\tilde{p}^\prime_{(n)}=p^\prime_{(n)},\quad\text{with}\;\; j\neq i,\\[0.5em]
\mathsf{S}_{\mathsf{4}}:&\quad \tilde{t}=-t,\quad
\tilde{x}^i=x^i,\quad \L\tilde{u}^i\R=-\L
u^i\R,\quad\tilde{p}=p,\quad\tilde{\nu}=-\nu,\nonumber\\
&\quad \tilde{u}^{\prime\hspace{0.25mm}
i}_{(n)}=-u^{\prime\hspace{0.25mm} i}_{(n)},\quad
\tilde{p}^\prime_{(n)}=p^\prime_{(n)},\label{161114:2111}
\end{align}
along with the two approximate symmetries, valid in a flow regime
where, respectively, the influence of viscosity is negligibly
small
\begin{align}
\mathsf{S}^{\mathsf{(approx.)}}_{\mathsf{5}}:&\quad\tilde{t}=t,\quad
\tilde{x}^i=e^{a}x^i,\quad\L\tilde{u}^i\R=e^{a}\L u^i\R,\quad
\L\tilde{p}\R=e^{2a}\L p\R,\nonumber\\
&\quad \tilde{u}^{\prime\hspace{0.25mm}
i}_{(n)}=e^{a}u^{\prime\hspace{0.25mm} i}_{(n)},\quad
\tilde{p}^\prime_{(n)}=e^{2a}p^\prime_{(n)},\label{161114:2112}
\end{align}
and/or the 3D turbulent flow is driven into a 2D
state:\footnote[2]{This confirms the result of \cite{Speziale98}
that the Reynolds stresses \eqref{161113:0838} are transforming
form-invariantly under all symmetries of the Navier-Stokes
equations. Note that although the Reynolds stresses
\eqref{161113:0838} do {\it not} transform as a tensor defined on
a change of coordinates for the two scaling symmetries
$\mathsf{S}_{\mathsf{1}}$ \eqref{161114:2119} and
$\mathsf{S}^{\mathsf{(approx.)}}_{\mathsf{5}}$~\eqref{161114:2112},
they nevertheless transform form-invariantly in the sense of
changing linearly and homogeneously.}
\begin{align}
\mathsf{S}^{\mathsf{(approx.)}}_{\mathsf{6}}:&\quad\tilde{t}=t,\quad
\tilde{x}^i=Q^i_{\,\, j}x^j,\quad\L\tilde{u}^i\R=Q^i_{\,\, j}\L
u^j\R+\dot{Q}^i_{\,\, j}x^j,\quad \tilde{u}^{\prime\hspace{0.25mm}
i}_{(n)}=Q^i_{\,\,j} u^{\prime\hspace{0.25mm} j}_{(n)},
\;\;\text{for}\;\; i,j=1,2,\nonumber\\
&\quad\text{and}\;\; \L\tilde{p}\R=\L
p\R+{\textstyle\frac{1}{2}}\omega_z^2\delta_{ij}x^ix^j+
2\omega_z\L\psi\R,\;\;\text{with}\;\; \L\psi\R=-\int_C\left(\L
u^2\R dx^1-\L u^1\R dx^2\right),\nonumber\\
&\quad\text{\phantom{and}}\;\;\tilde{p}^\prime_{(n)}=
p^\prime_{(n)}+ 2\omega_z \psi^\prime_{(n)},\;\;\text{with}\;\;
\psi^\prime_{(n)}=-\int_C\left(u^{\prime\hspace{0.25mm}2}_{(n)}
dx^1-u^{\prime\hspace{0.25mm}1}_{(n)}
dx^2\right).\label{161114:2103}
\end{align}
Now, the usual algebraic modelling ansatz for the Reynolds
stresses, as standardly favored in the turbulence community, is to
collectively choose them as an autonomous function of the mean
velocity field and its gradient only, not to include the mean
pressure field and the coordinates. Although there is no reason to
exclude the mean pressure and its gradient as modelling variables
\citep{Frewer09.2,Frewer09.3}, we will, for the sake of
simplicity, only proceed with the reduced, but still general
ansatz
\begin{equation}
\tau^{ij}=\tau^{ij}\Big(t,x^i,\nu\, ;\L u^i\R, \partial_i\L
u^j\R\Big).
\end{equation}
The aim is to model the Reynolds stresses such that they are
consistent with the Navier-Stokes equations, in particular to show
the same frame-indifference. Based on the results obtained in
Section~\ref{Sec3}, a self-evident proposal would be, for example,
\begin{multline}
\!\!(\nabla\cdot\vtau)^T=\phi_1\cdot\big(\hspace{0.25mm}\vx
-\vx_0^r\,\big)+\phi_2\cdot
\big(\hspace{0.25mm}\L\vu\R-\vu_0^r\,\big)+\phi_3\cdot
\left(\nabla\otimes \L\vu\R+\big(\nabla\otimes
\L\vu\R\big){}^T\,\right)
\cdot\big(\hspace{0.25mm}\vx-\vx_0^r\,\big)\\[0.25em]
+\phi_4\cdot\left(\big(\hspace{0.25mm}\L\vu\R-\vu_0^r\,\big){}^T\cdot
\nabla^T\right)
\L\vu\R+\phi_5\cdot\Delta\L\vu\R,\hspace{0.175cm}\label{161112:0921}
\end{multline}
already expressed for the gradient of the Reynolds stresses in a
3D covariant form. The $\phi_i$ are arbitrary scalar functions
with Euclidean-invariant arguments in all possible combinations
\begin{equation*}
\phi_i=\phi_i\Big(\nu,t-t_0^r,\lVert\vx-\vx_0^r\rVert,\lVert
\L\vu\R-\vu_0^r\rVert,\ldots,(\vx-\vx_0^r)^T\cdot(\L\vu\R-\vu_0^r),
(\vx-\vx_0^r)^T\cdot\Delta\L\vu\R,\ldots\Big),
\end{equation*}
but restricted to transform homogeneously in the following way
\begin{align}
\left.
\begin{array}{llllll}
\mathsf{S}_{\mathsf{1}}:&\;
\tilde{\phi}_1=e^{-4\varepsilon}\phi_1,&\;\;
\tilde{\phi}_2=e^{-2\varepsilon}\phi_2,&\;\;
\tilde{\phi}_3=e^{-2\varepsilon}\phi_3,&\;\;
\tilde{\phi}_4=\phi_4,&\;\; \tilde{\phi}_5=\phi_5,
\\[0.5em]
\mathsf{S}_{\mathsf{3}}:&\; \tilde{\phi}_1=\phi_1,&\;\;
\tilde{\phi}_2=\phi_2,&\;\;
\tilde{\phi}_3=\phi_3,&\;\;\tilde{\phi}_4=\phi_4,&\;\;
\tilde{\phi}_5=\phi_5,
\\[0.5em]
\mathsf{S}_{\mathsf{4}}:&\; \tilde{\phi}_1=\phi_1,&\;\;
\tilde{\phi}_2=-\phi_2,&\;\;\tilde{\phi}_3=-\phi_3,&\;\;
\tilde{\phi}_4=\phi_4,&\;\; \tilde{\phi}_5=-\phi_5,
\end{array}
~~~ \right\}
\end{align}
in order to guarantee frame-indifference under the transformations
\eqref{1611142101}-\eqref{161114:2111}. Note again that only
relative quantities enter the constitutive relation
\eqref{161112:0921}, where $t_0^r$, $\vx_0^r$ and $\vu_0^r$ are
some fixed absolute reference values characterizing the
environment of the system in which the fluid flows. As explained
in Section~\ref{Sec3}, these absolute system values transform in
the same way as the\linebreak respective variables characterizing
the flow itself for some constant value, that is, in the same way
as if the time coordinate $t$, the spatial coordinate $\vx$ and
the mean velocity $\L\vu\R$ were constant, independent of whether
$t_0^r$, $\vx_0^r$ and $\vu_0^r$ are specified in one particular
frame as zero or~not. Moreover, when also including the remaining
limit of the two approximative symmetries \eqref{161114:2112} and
\eqref{161114:2103}, the scalar functions in the constitutive
relation \eqref{161112:0921} are restricted further in that they
are to be modelled such that also the following limiting behavior
emerges
\begin{align}
\left.
\begin{array}{lllllll}
\mathsf{S}^{\mathsf{(approx.)}}_{\mathsf{5}}(\nu\rightarrow 0):&\;
\tilde{\phi}_1=\phi_1,&\; \tilde{\phi}_2=\phi_2,&\;
\tilde{\phi}_3=\phi_3,&\; \tilde{\phi}_4=\phi_4,&\;
\tilde{\phi}_5=e^{2a}\phi_5,
\\[0.5em]
\mathsf{S}^{\mathsf{(approx.)}}_{\mathsf{6}}\text{(3D}\rightarrow\text{2D)\color{blue}\footnotemark[3]}:&\;
\phi_2\rightarrow 0,&\; \phi_4\rightarrow 0. & & &
\end{array}
~~~\right\}\label{161114:2135}
\end{align}
\footnotetext[3]{The dependence on the velocity difference
$\L\vu\R-\vu_0^r$ has to vanish also in the contributing scalars
$\phi_1$, $\phi_3$ and~$\phi_5$.}\noindent Hence, in contrast to
traditional (algebraic) turbulence models for the Reynolds
stresses, the\linebreak present\hfill model\hfill
\eqref{161112:0921}\hfill exhibits\hfill an\hfill explicit\hfill
dependence\hfill on\hfill the\hfill mean\hfill velocity\hfill
field\hfill $\L\vu\R$,\hfill however,\hfill not
\newpage\noindent absolutely, but only in a relative manner, and
particularly such that in the instantaneous or fluctuating 2D
limit \eqref{161114:2135} these velocity contributions via
$\phi_2$ and $\phi_4$ will vanish. This is a novel modelling
feature which would be worth to investigate further in a future
study.

To close this section and this study, it is to mention that as one
prolongs from classical or ordinary thermodynamics to extended
thermodynamics by identifying the unclosed flux as a density
satisfying its own balance equation (see Appendix~\ref{SecD}), one
can analogously prolong from the mean momentum equations
\eqref{161111:1256} to the Reynolds stress transport equations by
identifying the Reynolds stress fluxes $\vtau$ as a set of
densities, to then obtain the extended system (in 3D covariant
form)
\begin{gather}
\left.
\begin{aligned}
\nabla\cdot\L \vu\R=0,\hspace{2.8cm}\\[0.5em]
\partial_t
\L\vu\R+\L\vu\R\cdot\nabla\L\vu\R+\nabla\cdot\vtau=-\nabla\L
p\R+\nu\Delta\L\vu\R,\\[0.6em]
\partial_t\vtau+\L \vu\R\cdot\nabla\vtau+\nabla\cdot
\vT=\nu\Delta\vtau+\vS,\hspace{0.9cm}
\end{aligned}
~~~\right\}\label{161112:2000}
\end{gather}
but, of course, at the price of gaining new unclosed terms: The
extended flux $\vT$ known as the turbulent transport driven by
velocity and pressure fluctuations, and the turbulent source
term~$\vS$ containing both positive (productive) and negative
(dissipative) contributions. In the spirit of extended
thermodynamics one can now assume for the unclosed terms solely a
local dependence on the densities $\L\vu\R$ and $\vtau$, including
a possible dependence on the space-time coordinates, but excluding
gradients and the pressure field $\L p\R$, which in the
incompressible case cannot be identified as an own independent
density evolving in time. A general ansatz for the two
constitutive relations would thus have the form$\,$\footnote[2]{In
how far this local ansatz inspired from extended thermodynamics is
sufficient to model turbulence is a different problem and has to
be investigated in a separate study. Maybe this ansatz is not
sufficient and the field gradients together with mean pressure
field have to be taken along as additional modelling variables.}
\begin{equation}
\vT=\vT(\nu,t,\vx,\L\vu\R,\vtau),\qquad\vS=\vS(\nu,t,\vx,\L\vu\R,\vtau).\label{161112:1917}
\end{equation}
The aim again is to model these relations such that they are
consistent with the Navier-Stokes equations, in particular to show
the same frame-indifference, that is, to be in accordance with the
Navier-Stokes symmetries \eqref{1611142101}-\eqref{161114:2103}.
Although a detailed analysis of this problem would be beyond the
scope of this section, one can nevertheless expect again an
explicit dependence on the mean velocity field, however, as before
in \eqref{161112:0921}, only as a velocity difference; and the
same for the coordinates if such a dependence is necessary to take
into account:\footnote[3]{The dependence on the Reynolds stresses
$\vtau$ need not to be considered as a difference, since $\vtau$
itself is already transforming form-invariantly under all
symmetries \eqref{1611142101}-\eqref{161114:2103} of the
Navier-Stokes equations.}
\begin{equation}
\vT=\vT(\nu,t-t_0^r,\vx-\vx_0^r,\L\vu\R-\vu_0^r,\vtau),
\qquad\vS=\vS(\nu,t-t_0^r,\vx-\vx_0^r,\L\vu\R-\vu_0^r,\vtau).\label{161113:0746}
\end{equation}
Hence, similar as the algebraic relation \eqref{161112:0921}, the
extended constitutive relations \eqref{161113:0746} will be
frame-indifferent only for those transformations
\eqref{1611142101}-\eqref{161114:2103} which are admitted as
symmetries or approximative symmetries by the instantaneous
Navier-Stokes equations. For all other transformations, like e.g.
a 3D uniform (time-dependent) rotation, the constitutive relations
\eqref{161112:0921} and \eqref{161113:0746} will no longer be
frame-indifferent relations anymore, as they will pick up frame
dependent terms $\vGamma$ in the sense as explained in
Section~\ref{Sec4.2}. In other words, when generally formulating
or transforming the balance equations \eqref{161112:2000} to
non-inertial systems, the corresponding constitutive relations
\eqref{161113:0746} will inherently show an explicit
frame-dependence of the form
\begin{equation}
\vT=\vT(\nu,t-t_0^r,\vx-\vx_0^r,\L\vu\R-\vu_0^r,\vtau;\vGamma),
\qquad\vS=\vS(\nu,t-t_0^r,\vx-\vx_0^r,\L\vu\R-\vu_0^r,\vtau;\vGamma).\label{161112:2003}
\end{equation}
Hence, modelling the constitutive relations \eqref{161112:2003}
generally frame-indifferent for {\it all} non-inertial systems
would therefore be incorrect; a problem faced, for example, in the
work of \cite{Dafalias07,Dafalias09,Dafalias11}, where in
particular the pressure-strain-rate correlations in~$\vS$ are
modelled explicitly frame-indifferent for all Euclidean
transformations, including the 3D time-dependent rotations.

\appendix
\titleformat{\section}
{\large\bfseries}{Appendix \thetitle.}{0.5em}{}
\numberwithin{equation}{section}

\newpage
\section{On the notion of geometrical invariance\label{SecA}}

The aim of this section is to show that not every geometric object
is automatically geometrically invariant when observed from
several different frames of reference: A geometric object which
stays invariant when observed from two different frames, may not
necessarily stay invariant when observed from another, a third
different frame. For example, let us consider as a geometric
object first a single point $\vx$ of physical 3D space, which we
observe from a fixed but arbitrarily chosen frame of reference
$\mathcal{F}$ spanned by a basis
$(\vg_1,\vg_2,\vg_3)$.\footnote[2]{For reasons of simplicity we
will only consider frame representations relative to a covariant
basis. To distinguish in this section in each frame between a
covariant basis $\vg_i$ and a contravariant basis $\vg^i$  would
only unnecessarily complicate the analysis, without gaining new
insight.} The geometric point $\vx$ in $\mathcal{F}$ then has the
unique representation
\begin{equation}
\vx=x^i \vg_i,\label{161028:1529}
\end{equation}
where $x^i$ are the three components of the point. It is clear
that both the components $x^i$ as well as the basis $\vg_i$ are
observer dependent and thus relative quantities which are assigned
only to the specific frame used, here currently to $\mathcal{F}$.
Now, let us observe this geometric point $\vx$ from a second,
different frame $\hat{\mathcal{F}}$ spanned by a basis
$(\hat{\vg}_1,\hat{\vg}_2,\hat{\vg}_3)$, and let the components in
each frame be related by a constant homogeneous linear
transformation (with an invertible matrix $\vA$)
\begin{equation}
\hat{x}^i=(\vA)^{i}_{\: j}\, x^j,\label{161028:1507}
\end{equation}
then it is clear that the geometric point $\vx$ itself is
unaffected from the process of which frame it is being observed,
since with \eqref{161028:1507} we only perform a relabelling of
that point.\footnote[3]{A transformation as \eqref{161028:1507},
connecting two different frames observing the same physical
object, i.e., two observers one object, is also known as a passive
transformation; opposed to an active transformation where the
observed object itself is transformed.} Hence, under the change of
frame $\mathcal{F}\rightarrow\hat{\mathcal{F}}$, according to
\eqref{161028:1507}, the point $\vx$ \eqref{161028:1529} stays
geometrically invariant
\begin{equation}
\vx=x^i\vg_i=\hat{x}^i\hat{\vg}_i.\label{161028:1750}
\end{equation}
This invariance then defines the transformation rule for the basis
as
\begin{equation}
\hat{\vg}_i=(\vA^{-T})_i^{\,\,\hspace{0.25mm}
j}\vg_j,\label{161028:1719}
\end{equation}
where $\vA^{-T}:=(\vA^{-1})^T$ is the short-hand notation for the
transpose of the inverse matrix of $\vA$. Now, let us consider
relative to the initial frame $\mathcal{F}$ a third frame
$\mathcal{F}^\prime$, which is spanned by the basis
$(\vg^\prime_1,\vg^\prime_2,\vg^\prime_3)$ and where the
components of the geometric point $\vx$ are now just related by a
constant non-zero shift
\begin{equation}
x^{\prime\, i}=x^i+b^i.\label{161028:1716}
\end{equation}
Since this transformation
$\mathcal{F}\rightarrow\mathcal{F}^\prime$ \eqref{161028:1716}
only resembles a parallel frame shift, it is clear that the new
basis $\vg_i^\prime$ is equivalent to the initial basis $\vg_i$,
i.e., $\vg_i^\prime=\vg_i$, $\forall i$. Thus, in contrast to the
first (homogeneous) linear transformation \eqref{161028:1507}, the
second (inhomogeneous) linear transformation~\eqref{161028:1716}
does not leave the considered geometric point $\vx$ invariant as
in \eqref{161028:1750}, since
\begin{equation}
\vx=x^i\vg_i=(x^{\prime\, i}-b^i)\vg^\prime_i=x^{\prime\,
i}\vg_i^\prime-b^i\vg_i^\prime\:\neq\: x^{\prime\, i}\vg_i^\prime
.
\end{equation}
Instead of absolute points, as $\vx$, only relative connections as
$\vr:=\vx_2-\vx_1$ between two absolute points $\vx_1$ and $\vx_2$
can serve as geometric invariants under \eqref{161028:1716}, since
only then obviously the invariance-breaking shift term cancels
\begin{equation}
\vr=(x^i_2-x^i_1)\vg_i=(x^{\prime\, i}_2-x^{\prime\,
i}_1)\vg^\prime_i.
\end{equation}
Also under the first linear transformation
$\mathcal{F}\rightarrow\hat{\mathcal{F}}$ \eqref{161028:1507}, the
relative connection $\vr$ remains to be a geometrical invariant
\begin{equation}
\vr=(x^i_2-x^i_1)\vg_i=(\hat{x}^i_2-\hat{x}^i_1)\hat{\vg}_i,
\end{equation}
as can be readily verified, by applying \eqref{161028:1719} and
\eqref{161028:1507} for each absolute point $\vx_1$ and $\vx_2$
separately. Hence, while the relative connection $\vr$ behaves
geometrically invariant under all three frames $\mathcal{F}$,
$\hat{\mathcal{F}}$ and $\mathcal{F}^\prime$ considered so far,
the absolute geometric point $\vx$ is only invariant when observed
from $\mathcal{F}$ and $\hat{\mathcal{F}}$, but not when observed
from  $\mathcal{F}^\prime$. Although this example already suffices
to show that a geometric object need not to be necessarily
invariant when observed from different frames of reference, it is
expedient to continue this geometric process of
invariance-breaking to gain further insights. For example, the
frame to be considered next will break the geometrical invariance
of the finite difference vector $\vr$.

Let us now consider, relative to the initial frame $\mathcal{F}$,
a fourth frame $\mathcal{F}^*$, spanned by a basis
$(\vg^*_1,\vg^*_2,\vg^*_3)$ which is local in space, i.e.,
$\vg^*_i=\vg^*_i(\vx)$, $\forall i$, and where the components of
any geometric point $\vx$ in each frame are related by an
arbitrary and locally invertible spatial coordinate transformation
\begin{equation}
x^{*\, i}=x^{*\, i}(x^j).\label{161028:2207}
\end{equation}
Since the new frame $\mathcal{F}^*$ can be identified as a linear
space locally in each point $\vx$, it is clear that the new basis
$(\vg^*_1,\vg^*_2,\vg^*_3)$ can always be expressed as a linear
combination of the initial basis $(\vg_1,\vg_2,\vg_3)$ locally in
each point of the initial linear space $\mathcal{F}$ --- similar
to \eqref{161028:1719}, however only locally in each (geometric)
point
\begin{equation}
\vg^*_i(\vx)=\big(\vJ^{-T}(\vx)\big)_i^{\,\,
j}\vg_j.\label{161028:2040}
\end{equation}
Obviously the finite difference vector $\vr$ cannot be observed as
a geometrical invariant from the fourth and local frame
$\mathcal{F}^*$, since for each point one faces the non-invariant
and meaningless relation
\begin{equation}
\vr=(x^i_2-x^i_1)\vg_i=x_2^i\,\big(\vJ^T(\vx_2)\big)_i^{\,\,
j}\vg^*_j(\vx_2)\Big|_{x_2^{ k}=x_2^{k}(x_2^{*
l})}-x_1^i\,\big(\vJ^T(\vx_1)\big)_i^{\,\,
j}\vg^*_j(\vx_1)\Big|_{x_1^{ k}=x_1^{k}(x_1^{* l})}.
\end{equation}
A meaningful transformation relation can only be obtained in the
infinitesimal limit $\vx_2\rightarrow\vx_1$, i.e., only for an
infinitesimal difference vector $d\vr$, which in the local frame
$\mathcal{F}^*$ now even turns out to be geometrically invariant,
since according to \eqref{161028:2207} and \eqref{161028:2040} we
obtain the reduction
\begin{equation}
d\vr=dx^i\vg_i=\frac{\partial x^i(x^{*k})}{\partial x^{* j}} dx^{*
j}\big(\vJ^T(\vx)\big)_i^{\,\, j}\vg^*_j(\vx)=dx^{*
j}\vg^*_j(\vx)\Big|_{x^{k}=x^{k}(x^{*l})}=:dx^{*i}\vg^*_{i},
\end{equation}
if the local linear mapping $\vJ$ \eqref{161028:2040} is
identified as the Jacobi matrix
\begin{equation}
\big(\vJ(\vx)\big)^i_{\,\, j}=\frac{\partial x^{*i}(x^k)}{\partial
x^{j}}.
\end{equation}
Since the two linear transformation \eqref{161028:1507} and
\eqref{161028:1716} are both contained as special transformations
of \eqref{161028:2207}, we can conclude that the geometrical
object $d\vr$ is invariant under all four frames $\mathcal{F}$,
$\hat{\mathcal{F}}$, $\mathcal{F}^\prime$ and $\mathcal{F}^*$
considered so far, while the invariance of $\vr$ between the three
former frames is broken for the later frame $\mathcal{F}^*$.

As the last example, we now will consider a frame which also
breaks the geometrical invariance of $d\vr$. Relative again to the
initial frame $\mathcal{F}$, let us consider a fifth frame
$\tilde{\mathcal{F}}$ spanned by a basis
$(\tilde{\vg}_1,\tilde{\vg}_2,\tilde{\vg}_3)$ where the components
of a geometric point $\vx$ in each frame are again related by the
homogenous linear transformation \eqref{161028:1507}, but now
through a time-dependent and thus non-constant matrix
$\vA=\vA(t)$. This dependence on a fourth coordinate breaks the
invariant property of the infinitesimal spatial 3D-vector $d\vr$,
since the total derivative of the underlying transformation for
the coordinates connecting the frames $\mathcal{F}$ and
$\tilde{\mathcal{F}}$
\begin{equation}
\tilde{x}^i=\tilde{x}^i(x^j,t)=\big(\vA(t)\big)^{i}_{\: j}\,
x^j,\;\text{with its local inverse:}\;\;
x^i=x^i(\tilde{x}^j,t)=\big(\vA^{-1}(t)\big)^{i}_{\: j}\,
\tilde{x}^j,\label{161028:2342}
\end{equation}
is now given as
\begin{equation}
d\tilde{x}^i=\frac{\partial\tilde{x}^i}{\partial x^j}
dx^j+\frac{\partial\tilde{x}^i}{\partial
t}dt=\big(\vA(t)\big)^{i}_{\: j}\, dx^j
+\big(\dot{\vA}(t)\big)^{i}_{\: j}\, x^j\, dt,\label{161029:1532}
\end{equation}
which thus, due to respecting the variation also along the time
coordinate $t$, leads to the non-invariant relation
\begin{align}
d\vr=dx^i\vg_i &=\left(\big(\vA^{-1}(t)\big)^{i}_{\: j}\,
d\tilde{x}^j
+\big({\textstyle\frac{d}{dt}}\vA^{-1}(t)\big)^{i}_{\: j}\,
\tilde{x}^j\, dt\right)\big(\vA^T(t)\big)_i^{\,\, k}\tilde{\vg}_k
\nonumber\\[0.5em]
&=\, d\tilde{x}^j\tilde{\vg}_j+\big(\vA(t){\textstyle\frac{d}{dt}}
\vA^{-1}(t)\big)^{k}_{\,\,\hspace{0.25mm} j}\, \tilde{x}^j
dt\,\tilde{\vg}_k \:\neq\:
d\tilde{x}^i\tilde{\vg}_i,\label{161029:1311}
\end{align}
where we used the obvious fact that the basis still transforms as
formulated in \eqref{161028:1719}, however, in the current case,
locally now for each time step.

A geometrical invariance for $d\vr$ under time-dependent
transformations can only achieved when changing the framework from
3D space to a 4D space-time. If the aim is to remain within the
realm of classical Newtonian physics then it is necessary to only
consider space-time transformations in which the time coordinate
transforms absolutely, i.e. invariantly up to some constant time
shift. For that, let us reformulate the two frames $\mathcal{F}$
and $\tilde{\mathcal{F}}$ currently considered as Newtonian frames
in a true 4D space-time setting, $\mathcal{F}^{\,\text{4D}}$ and
$\tilde{\mathcal{F}}^{\,\text{4D}}$, spanned now by a 4D basis
$(\vg_0,\vg_1,\vg_2,\vg_3)$ and
$(\tilde{\vg}_0,\tilde{\vg}_1,\tilde{\vg}_2,\tilde{\vg}_3)$,
respectively. The components of a geometric space-time
point~$\vx^{\text{4D}}$ in each frame $\mathcal{F}^{\,\text{4D}}$
and $\tilde{\mathcal{F}}^{\,\text{4D}}$ are again related by
\eqref{161028:2342}, however, now reformulated into a 4D Newtonian
framework:\footnote[2]{The Greek indices run from $0$ to $3$, the
Latin indices from $1$ to $3$, and $x^0=t$ is defined as the time
coordinate.\linebreak In \cite{Frewer09.1,Frewer09.2,Frewer09.3}
it is demonstrated that the metric of the underlying 4D manifold
allowing for transformations of the kind \eqref{161029:1205}
degenerates into a singular twofold metric, namely into a
time-like and into a space-like metric.}
\begin{equation}
\tilde{x}^\alpha=\tilde{x}^\alpha(x^\beta),\;\;\text{where}\;\;
\tilde{x}^0=x^0,\;\text{and}\;\;
\tilde{x}^i=\big(\vA(x^0)\big)^{i}_{\: j}\,
x^j.\label{161029:1205}
\end{equation}
The infinitesimal 4D-vector $d\vr^{\text{4D}}$ transforms
geometrically invariant
\begin{equation}
d\vr^{\text{4D}}=dx^\alpha \vg_\alpha=\frac{\partial
x^\alpha}{\partial\tilde{x}^\beta}d\tilde{x}^\beta \frac{\partial
\tilde{x}^\gamma}{\partial
x^\alpha}\tilde{\vg}_\gamma=\delta^{\gamma}_\beta d\tilde{x}^\beta
\tilde{\vg}_\gamma=d\tilde{x}^\alpha \tilde{\vg}_\alpha,
\end{equation}
since the basis $\vg_0$ for the time coordinate compensates the
invariance-breaking term \eqref{161029:1311} in the pure 3D case.
In particular, the basis $(\vg_\alpha)=(\vg_0,\vg_i)$ transforms
as
\begin{equation}
\left.
\begin{aligned}
\tilde{\vg}_0&=\frac{\partial
x^\alpha}{\partial\tilde{x}^0}\vg_\alpha=\vg_0+\frac{\partial
x^i}{\partial\tilde{x}^0}\vg_i=\vg_0+\big({\textstyle\frac{d}{dt}}\vA^{-1}(t)\big)^{i}_{\:
j}\tilde{x}^j\vg_i,\\[0.5em]
\tilde{\vg}_i&=\frac{\partial
x^\alpha}{\partial\tilde{x}^i}\vg_\alpha=\frac{\partial
x^j}{\partial\tilde{x}^i}\vg_j
=\big(\vA^{-T}(t)\big)_{i}^{\,\,\hspace{0.25mm} j}\vg_j.
\end{aligned}
~~~~~\right\}
\end{equation}
This completes the investigation on geometrical invariance, to
show that it depends on the frame used whether a geometrical
object is observed invariantly or not. Worthwhile to note in this
respect is that the notion of geometrical invariance is intimately
linked to form-invariance and the definition of a tensor (see
Sections \ref{Sec1} and \ref{Sec2}): For example, in the frame
relation $\mathcal{F}\leftrightarrow\hat{\mathcal{F}}$ the
absolute space point $\vx$ is observed as a geometric invariant
because the coordinates \eqref{161028:1507} transform as a
(contravariant) tensor of rank 1, while in the frame relation
$\mathcal{F}\leftrightarrow\mathcal{F}^\prime$ this point is not
observed as a invariant, because in this case the coordinates
\eqref{161028:1716} do not transform as a tensor. Or, for example,
$d\vr$ which is observed as a geometrical invariant in
$\mathcal{F}\leftrightarrow\mathcal{F}^*$, but not in
$\mathcal{F}\leftrightarrow\tilde{\mathcal{F}}$, simply because in
this case again the coordinates \eqref{161029:1532} do not
transform as a tensor,\linebreak but only in a 3D formulation,
while in true 4D formulation they do.

\section{The non-tensors for uniform rotations in a 3D formulation\label{SecB}}

The non-tensors considered in this study are the velocity field
$\vu(\vx)$, its gradient $\vL(\vx)=\nabla\otimes \vu(\vx)$, and
the vorticity
$\vW(\vL(\vx))=\frac{1}{2}\big(\vL(\vx)-\vL(\vx)^T\big)$. Within a
true 4D framework, however, all three quantities naturally appear
as tensors \citep{Frewer09.1}, but not in a 3D framework as
formulated herein. Nevertheless, it is indeed possible to turn
these three quantities into tensors without changing to a 4D
framework, namely by just making use of a similar procedure from
differential geometry when defining an affine connection to
overcome the general non-tensor property of the second and
higher-order derivatives; ultimately thus supporting once more the
objection of Kretschmann (see discussion in Section \ref{Sec4.1}).
To demonstrate this, consider first the central non-tensor in this
list, the velocity field $\vu(\vx)$, which transforms as
\eqref{161020:1425}
\begin{equation}
\tilde{\vu}(\tilde{\vx})+\vOmega\tilde{\vx}=\vQ\vu(\vx),\label{161018:1229}
\end{equation}
where again it is clear that the inhomogeneous term proportional
to the spin $\vOmega$ destroys the tensor property of the velocity
field in 3D
--- if everything, however, would be reformulated
into a true 4D framework (within Newtonian physics), this
`non-tensor'-problem would not exist in the first place
\citep{Frewer09.1,Frewer09.2}, since the 4D velocity field
$u^\alpha(x^\gamma)=(1,\vu(x^0,\vx))$ would then naturally always
transform as a tensor$\,$\footnote[2]{Note that the Newtonian 4D
velocity field $u^\alpha(x^\gamma)=(1,\vu(x^0,\vx))$ is a tensor
that never gets zero, i.e.~it is non-zero in all reference frames
and thus does not distinguish any particular frame, a feature
inherent to the definition of a tensor (see footnote on
p.$\,$\pageref{N1}). Finally note here again that all Greek
indices run from $0$~to~$3$, the Latin indices from $1$ to $3$,
and $x^0=t$ is defined as the time coordinate, which itself stays
invariant under the spatial coordinate transformation
\eqref{161016:1904}, i.e. $\tilde{x}^0=x^0$.}
\begin{equation}
\tilde{u}^\alpha(\tilde{x}^\gamma)=\frac{\partial\tilde{x}^\alpha}{\partial
x^\beta}u^\beta(x^\gamma),\;\,\text{where}\;\;
\frac{\partial\tilde{x}^i}{\partial x^j}=(\vQ)^i_{\;\, j},\;\;
\frac{\partial\tilde{x}^i}{\partial x^0}=(\dot{\vQ})^i_{\;\,
j}x^j,\;\,\text{and}\;\;\frac{\partial \tilde{x}^0}{\partial
x^j}=0,\;\; \frac{\partial\tilde{x}^0}{\partial x^0}=1.\;\,
\end{equation}
But our aim in this study is to reside within 3D and not to change
the framework to 4D. Hence a different argument has to be used
when trying to reformulate \eqref{161018:1229} into a 3D tensor
relation. For that a short excursion into differential geometry is
necessary, by recalling the fact that also in a true 4D framework
one naturally runs into non-tensors. The most natural one is the
derivative of any covariant vector field
$A_\alpha=A_\alpha(x^\gamma)$, which for {\it arbitrary}
space-time coordinate transformations changes as
\citep{Frewer09.1}
\begin{align}
\frac{\partial\tilde{A}_\alpha}{\partial \tilde{x}^\beta}
&=\frac{\partial}{\partial \tilde{x}^\beta}\left(\frac{\partial
x^\rho}{\partial \tilde{x}^\alpha}A_\rho\right)=\frac{\partial
x^\rho}{\partial \tilde{x}^\alpha}\frac{\partial}{\partial
\tilde{x}^\beta}A_\rho + \frac{\partial^2 x^\rho}{\partial
\tilde{x}^\alpha\partial\tilde{x}^\beta}A_\rho\nonumber\\[0.25em]
&=\frac{\partial x^\rho}{\partial \tilde{x}^\alpha}\frac{\partial
x^\sigma}{\partial \tilde{x}^\beta}\frac{\partial A_\rho}{\partial
x^\sigma} + \frac{\partial^2 x^\rho}{\partial
\tilde{x}^\alpha\partial\tilde{x}^\beta}A_\rho \;\,\neq\;\,
\frac{\partial x^\rho}{\partial \tilde{x}^\alpha}\frac{\partial
x^\sigma}{\partial \tilde{x}^\beta}\frac{\partial A_\rho}{\partial
x^\sigma},\label{161017:1508}
\end{align}
where for a better readability the explicit functional coordinate
dependence was dropped. As a result, we see that the second
derivative term as an inhomogeneous term destroys the tensor
property, similar as the spin term $\vOmega$ in
\eqref{161018:1229}. However, by redefining expressions, it is
possible to turn \eqref{161017:1508} in a tensor relation. The
usual procedure, as taken from \cite{Schroedinger50}, is first to
introduce the short-hand notation
\begin{equation}
\tilde{\Gamma}^\mu_{\alpha\beta}:=\frac{\partial\tilde{x}^\mu}{\partial
x^\rho}\frac{\partial^2 x^\rho}{\partial
\tilde{x}^\alpha\partial\tilde{x}^\beta}\neq 0,
\end{equation}
and then to rewrite relation \eqref{161017:1508} equivalently~as
\begin{equation}
\frac{\partial\tilde{A}_\alpha}{\partial \tilde{x}^\beta}-
\tilde{\Gamma}^\mu_{\alpha\beta}\tilde{A}_\mu =\frac{\partial
x^\rho}{\partial \tilde{x}^\alpha}\frac{\partial
x^\sigma}{\partial \tilde{x}^\beta}\left(\frac{\partial
A_\rho}{\partial
x^\sigma}-\Gamma^\mu_{\rho\sigma}A_\mu\right),\;\;
\text{where}\;\; \Gamma^\mu_{\rho\sigma}=0,\label{161017:1818}
\end{equation}
which allows us to see that the formal quantity$\,$\footnote[3]{In
the literature the expression $\nabla_\sigma A_\rho$ is sometimes
also denoted as $A_{\rho;\sigma}$, opposed to the usual partial
derivative denoted as $\partial_\sigma A_\rho=A_{\rho,\sigma}$.}
\begin{equation}
\nabla_\sigma A_\rho:=\frac{\partial A_\rho}{\partial
x^\sigma}-\Gamma^\mu_{\rho\sigma}A_\mu,\label{161121:1243}
\end{equation}
transforms as a covariant tensor of rank 2
\begin{equation}
\tilde{\nabla}_\beta\tilde{A}_\alpha=\frac{\partial
x^\rho}{\partial \tilde{x}^\alpha}\frac{\partial
x^\sigma}{\partial \tilde{x}^\beta}\nabla_\sigma
A_\rho,\label{161017:1806}
\end{equation}
where $\nabla_\alpha$ is known as the covariant derivative
satisfying all linear rules of ordinary differentiation as
$\partial_\alpha$. Although we managed with \eqref{161017:1806} to
define a form-invariant differentiation process from the
non-tensor relation \eqref{161017:1508}, the explicit frame
dependence could not be removed and still is to be found in the
transformed quantity
$\tilde{\nabla}_\beta\tilde{A}_\alpha=\tilde{\partial}_\beta\tilde{A}_\alpha-
\tilde{\Gamma}^\mu_{\alpha\beta}\tilde{A}_\mu$, being, due to
$\tilde{\Gamma}^\mu_{\alpha\beta}$, a frame-dependent
(non-objective) quantity, except in the special case when
considering linear space-time coordinate transformations
connecting inertial frames of reference, which, within classical
Newtonian physics, are the Galilei transformations. Otherwise, for
all other (non-inertial) frames of references, the transformed
covariant derivative $\tilde{\nabla}_\beta\tilde{A}_\alpha$ shows
a regular dependence on the used frame through its driving and
defining constituent $\tilde{\Gamma}^\mu_{\alpha\beta}$, which
itself transforms as a non-tensor again. To determine this
transformation it is necessary to identify \eqref{161017:1818} as
an equation for $\tilde{\Gamma}^\mu_{\alpha\beta}$ and then to
solve it in terms of $\Gamma^\mu_{\rho\sigma}$, which can be
achieved by rewriting \eqref{161017:1818} equivalently as
\begin{align}
- \tilde{\Gamma}^\mu_{\alpha\beta}\tilde{A}_\mu
&=-\frac{\partial\tilde{A}_\alpha}{\partial
\tilde{x}^\beta}+\frac{\partial x^\rho}{\partial
\tilde{x}^\alpha}\frac{\partial x^\sigma}{\partial
\tilde{x}^\beta}\left(\frac{\partial A_\rho}{\partial
x^\sigma}-\Gamma^\mu_{\rho\sigma}A_\mu\right)\nonumber\\[0.5em]
&=-\left(\frac{\partial x^\rho}{\partial
\tilde{x}^\alpha}\frac{\partial x^\sigma}{\partial
\tilde{x}^\beta}\frac{\partial A_\rho}{\partial x^\sigma} +
\frac{\partial^2 x^\rho}{\partial
\tilde{x}^\alpha\partial\tilde{x}^\beta}A_\rho\right)+\frac{\partial
x^\rho}{\partial \tilde{x}^\alpha}\frac{\partial
x^\sigma}{\partial \tilde{x}^\beta}\left(\frac{\partial
A_\rho}{\partial
x^\sigma}-\Gamma^\mu_{\rho\sigma}A_\mu\right)\nonumber\\[0.5em]
&=-\frac{\partial^2 x^\rho}{\partial
\tilde{x}^\alpha\partial\tilde{x}^\beta}A_\rho-\frac{\partial
x^\rho}{\partial \tilde{x}^\alpha}\frac{\partial
x^\sigma}{\partial \tilde{x}^\beta}\Gamma^\nu_{\rho\sigma}A_\nu
=-\left(\frac{\partial^2 x^\rho}{\partial
\tilde{x}^\alpha\partial\tilde{x}^\beta}\frac{\partial\tilde{x}^\mu}
{\partial x^\rho}+\frac{\partial x^\rho}{\partial
\tilde{x}^\alpha}\frac{\partial x^\sigma}{\partial
\tilde{x}^\beta}\frac{\partial\tilde{x}^\mu} {\partial
x^\nu}\Gamma^\nu_{\rho\sigma}\right)\tilde{A}_\mu,\;\,
\end{align}
from which then the solution can be read off as (since this
relation must hold for all $\tilde{A}_\mu$)
\begin{equation}
\tilde{\Gamma}^\mu_{\alpha\beta}=\frac{\partial\tilde{x}^\mu}
{\partial x^\nu}\frac{\partial x^\rho}{\partial
\tilde{x}^\alpha}\frac{\partial x^\sigma}{\partial
\tilde{x}^\beta}\Gamma^\nu_{\rho\sigma}+\frac{\partial\tilde{x}^\mu}
{\partial x^\rho}\frac{\partial^2 x^\rho}{\partial
\tilde{x}^\alpha\partial\tilde{x}^\beta}.\label{161019:0905}
\end{equation}
Thus $\Gamma^\mu_{\alpha\beta}$ transforms as a non-tensor, that
is, if it vanishes in one coordinate system
$\Gamma^\mu_{\alpha\beta}=0$ (initial coordinate system as used in
\eqref{161017:1818}), then it does not necessarily vanish in any
other coordinate system $\tilde{\Gamma}^\mu_{\alpha\beta}\neq0$.
Hence $\Gamma^\mu_{\alpha\beta}$ is a non-trivial quantity known
as the affine connection which can be arbitrarily assigned in one
coordinate system determining then in that considered system the
meaning of parallel displacement in space-time by connecting
nearby tangent spaces. To conclude, we thus managed to turn the
non-tensor relationship \eqref{161017:1508} into the
form-invariant, but still non-objective tensor relationship
\eqref{161017:1806}.

Obviously, the idea of the above procedure in defining an affine
connection can now also be applied to the 3D non-tensor relation
\eqref{161018:1229}. By equivalently rewriting this relation in
the way as it was done in \eqref{161017:1818}
\begin{equation}
\tilde{\vu}(\tilde{\vx})+\tilde{\vOmega}\tilde{\vx} =
\vQ\Big(\vu(\vx)+\vOmega\vx\Big),\;\;\text{where}\;\;
\tilde{\vOmega}:=\vQ\dot{\vQ}^T\neq
\boldsymbol{0},\;\text{and}\;\;\vOmega=\boldsymbol{0},\label{161018:0858}
\end{equation}
and then, similar as in \eqref{161017:1806}, redefining this
expression into the tensor relation
\begin{equation}
\tilde{\vu}_{\tilde{\vOmega}}(\tilde{\vx})=\vQ\vu_{\vOmega}
(\vx),\label{161018:0847}
\end{equation}
we finally yield a 3D form-invariant
representation$\,$\footnote[2]{Note that the specifically designed
3D-form-invariance \eqref{161018:0847} is not equivalent to the
form-invariance that would be achieved in a true 4D formulation
\citep{Frewer09.1,Frewer09.2,Frewer09.3}. Form-invariance of the
velocity field in 4D is of a different nature than
\eqref{161018:0847}. The former concept is absolute, while the
latter one is relative; see the definitions at the end of this
section.} of \eqref{161018:1229}, with
$\vu_{\vOmega}(\vx):=\vu(\vx)+\vOmega\vx$ being the new redefined
velocity field relative to the spin term $\vOmega$. Although being
manifestly form-invariant, the explicit frame-dependence
originating from \eqref{161018:1229}, however, could not be
removed in \eqref{161018:0847}: The transformed redefined velocity
$\tilde{\vu}_{\tilde{\vOmega}}(\tilde{\vx})$ still shows a regular
dependency on the uniform rotating frame through its spin
$\tilde{\vOmega}$, which itself transforms as a non-tensor.
Similar as in the procedure outlined above for the affine
connection~$\Gamma^\mu_{\alpha\beta}$, the transformation of
$\tilde{\vOmega}$ is determined from \eqref{161018:0858} by
equivalently rewriting it as
\begin{align}
\tilde{\vOmega}\tilde{\vx}
&=-\tilde{\vu}(\tilde{\vx})+\vQ\Big(\vu(\vx)+\vOmega\vx\Big)\nonumber\\[0.5em]
&=-\Big(\vQ\vu(\vx)+\dot{\vQ}\vx\Big)+\vQ\Big(\vu(\vx)+\vOmega\vx\Big)
=-\dot{\vQ}\vQ^T\tilde{\vx}+\vQ\vOmega\vQ^T\tilde{\vx},
\end{align}
and then to solve it for $\tilde{\vOmega}$ (valid for all
$\tilde{\vx}$) to obtain the (non-objective) transformation rule
\begin{equation}
\tilde{\vOmega}=\vQ\vOmega\vQ^T+\vQ\dot{\vQ}^T,\label{161019:0925}
\end{equation}
which can be formally compared to the situation in
\eqref{161019:0905}: The spin $\vOmega$ transforms as a
non-tensor, that is, if it vanishes in one coordinate system
$\vOmega=\boldsymbol{0}$ (initial coordinate system as used in
\eqref{161018:0858}), then it does not necessarily vanish in any
other coordinate system $\tilde{\vOmega}\neq\boldsymbol{0}$.
Hence, the spin $\vOmega$ can be identified as an own independent
quantity which can be arbitrarily assigned to feature one uniform
rotating system which then, via \eqref{161019:0925}, can be
connected to another uniform rotating system $\tilde{\vOmega}$.
The two different (uniform rotating) systems, tilde and non-tilde,
are then connected by a relative coordinate transformation
\begin{equation}
\tilde{\vx}=\vQ  \vx, \quad\text{with}\;\;
\vQ\vQ^T=\boldsymbol{1},\;\; \text{det}(\vQ)=1,
\quad\text{and}\;\; \vOmega_R := \vQ\dot{\vQ}^T, \;\;
\dot{\vOmega}_R=\boldsymbol{0},\label{161021:1159}
\end{equation}
having again, of course, the same property of an orientated
uniform rotation as initially defined in \eqref{161016:1904}, but,
with a relative spin $\vOmega_R$ which in general is different now
to the initially chosen (untransformed) spin $\vOmega$, which
itself can be assigned arbitrarily. In other words, from a more
constructive point of view, we thus face an initial uniform
rotating system, characterized by an arbitrarily assigned
$\vOmega$, that gets transformed by \eqref{161021:1159} to then
obtain a new uniform rotating frame characterized by
$\tilde{\vOmega}$, or, from a preset point of view, we already
face a configuration of two different uniform rotating frames
$\vOmega$ and $\tilde{\vOmega}$ which then are both connected by
the relative transformation \eqref{161021:1159} with spin
$\vOmega_R$. In both cases, if the first spin $\vOmega$ is chosen
arbitrarily, then the set of values for the second spin
$\tilde{\vOmega}$ are uniquely fixed by relation
\eqref{161019:0925} as
\begin{equation}
\tilde{\vOmega}=\vQ\vOmega\vQ^T+\vOmega_R, \;\;\text{for any
given}\;\, \vOmega_R = \vQ\dot{\vQ}^T.\label{161021:1226}
\end{equation}
To conclude, we thus managed to turn the non-tensor relationship
\eqref{161018:1229} into the form-invariant, but still
non-objective tensor relationship \eqref{161018:0847}. This
relation allows us now to also turn the non-tensor relationships
of the velocity gradient and the vorticity
\begin{equation}
\left.
\begin{aligned}
&\tilde{\vL}(\tilde{\vx})+\vOmega = \vQ\vL(\vx)\vQ^T,
\\[0.25em]
&\tilde{\vW}(\tilde{\vL}(\tilde{\vx}))+\vOmega =
\vQ\vW(\vL(\vx))\vQ^T,
\end{aligned}
~~~~~ \right\}\label{161019:0940}
\end{equation}
into their 3D-form-invariant, but still frame-dependent
(non-objective) tensor relations
\begin{equation}
\left.
\begin{aligned}
&\tilde{\vL}_{\tilde{\vOmega}}(\tilde{\vx})=
\vQ\vL_{\vOmega}(\vx)\vQ^T,
\\[0.25em]
&\tilde{\vW}(\tilde{\vL}_{\tilde{\vOmega}}(\tilde{\vx})) =
\vQ\vW(\vL_{\vOmega}(\vx))\vQ^T,
\end{aligned}
~~~~~~~ \right\}\label{161019:0943}
\end{equation}
with
$\vL_{\vOmega}(\vx):=\nabla\otimes\vu_{\vOmega}(\vx)=\vL(\vx)+\vOmega$
and $\vW(\vL_{\vOmega}(\vx)):=\frac{1}{2}\big(\vL_{\vOmega}(\vx)
-\vL_{\vOmega}(\vx)^T\big)$ being, relative to the spin term
$\vOmega$, the new redefined velocity gradient and vorticity,
respectively.

Despite the resemblance of redefining the generally transformed
derivative of a covariant vector field in 4D, on the one hand, and
the uniform rotated velocity field in 3D, on the other, as the
form-invariant tensor relation \eqref{161017:1806} and
\eqref{161018:0847}, respectively, there is nevertheless a
decisive difference between these two approaches: While the former
construction procedure can be\linebreak straightforwardly
generalized to derivatives of tensors of any kind, namely
as~\citep{Schroedinger50}
\begin{equation*}
\partial_\nu A^{\kappa\lambda\ldots}_{\rho\sigma\ldots}
\;\rightarrow\; \nabla_\nu
A^{\kappa\lambda\ldots}_{\rho\sigma\ldots} =\partial_\nu
A^{\kappa\lambda\ldots}_{\rho\sigma\ldots}+
\Gamma^\kappa_{\alpha\nu}A^{\alpha\lambda\ldots}_{\rho\sigma\ldots}+
\Gamma^\lambda_{\alpha\nu}A^{\kappa\alpha\ldots}_{\rho\sigma\ldots}
+\cdots -
\Gamma^\alpha_{\rho\nu}A^{\kappa\lambda\ldots}_{\alpha\sigma\ldots}
-\Gamma^\alpha_{\sigma\nu}A^{\kappa\lambda\ldots}_{\rho\alpha\ldots}-\cdots,
\end{equation*}
without knowing the explicit structure of the general tensor
$A^{\kappa\lambda\ldots}_{\rho\sigma\ldots}$ itself, one fails to
do so for the latter construction process if one would generalize
to some arbitrary non-tensor
\begin{equation}
\vN(\vx)\;\rightarrow\;
\vN_{\vOmega}(\vx)=\vN(\vx)+\vF\big(\vOmega,\vN,\{\vX_n\}\big),
\end{equation}
since, without knowing the explicit structure of $\vN$, it is not
clear beforehand how this quantity transforms under a uniform
rotation induced by $\vOmega$, i.e., the redefining function $\vF$
is not known beforehand unless the functional structure of $\vN$
is explicitly known --- the set of extra variables~$\vX_n$ in
$\vF$ denote all quantities that may arise when explicitly
transforming $\vN$ into a uniform rotating
frame.\footnote[2]{Moreover, it is to note that a redefinition
$\vN\rightarrow\vN_{\vOmega}=\vN+\vF$ is only valid if the
redefining function $\vF$ is also compatible with the
transformation rule \eqref{161019:0925} for $\vOmega$.} In other
words, the introduced process in 3D of redefining non-tenors to
tensors is context-related; in clear contrast to the similar
process considered in 4D when redefining the partial to a
covariant derivative. Hence, if a redefinition for
$\vOmega\neq\boldsymbol{0}$ is possible, we thus still have to
distinguish between those tensors which have the property
$\vN=\vN_{\vOmega}$, i.e., where $\vF=\boldsymbol{0}$, which
appropriately can be classified as {\it absolute} tensors or
simply as tensors, and those where $\vF\neq\boldsymbol{0}$ as {\it
relative} tensors. This classification is used in
Sections~\ref{Sec2.3}$\,$-$\,$\ref{Sec2.6}.

\section{The explicit frame-dependence of the
relative vorticity\label{SecC}}

As introduced in the previous Appendix \ref{SecB}, the relative
vorticity $\vW(\vL_{\vOmega}(\vx))$ transforms form-invariantly as
\eqref{161019:0943}
\begin{equation}
\tilde{\vW}(\tilde{\vL}_{\tilde{\vOmega}}(\tilde{\vx}))
=\vQ\vW(\vL_{\vOmega}(\vx))\vQ^T,\label{161021:1544}
\end{equation}
where $\tilde{\vOmega}$ and $\vOmega$ are two unequal spins
featuring two different uniform rotating (non-inertial) reference
frames which are connected through a coordinate transformation
$\tilde{\vx}=\vQ\vx$ \eqref{161021:1159} via the relation
\eqref{161021:1226}
\begin{equation}
\tilde{\vOmega}=\vQ\vOmega\vQ^T+\vOmega_R, \;\;\text{with}\;\,
\vOmega_R = \vQ\dot{\vQ}^T.\label{161021:1540}
\end{equation}
That the relative vorticity $\vW(\vL_{\vOmega}(\vx))$ constitutes
a frame-dependent (non-objective) quantity is manifestly clear,
since its evaluation
$\vW(\vL_{\vOmega}(\vx))=\vW(\vL(\vx))+\vOmega$ explicitly depends
on the spin value $\vOmega$ of the chosen frame
--- the same is of course also true for the relative vorticity
in the transformed domain
$\tilde{\vW}(\tilde{\vL}_{\tilde{\vOmega}}(\tilde{\vx}))
=\tilde{\vW}(\tilde{\vL}(\tilde{\vx}))+\tilde{\vOmega}$, which
explicitly depends, via \eqref{161021:1540}, on the spin value
$\tilde{\vOmega}$  of the second frame. Hence the correct approach
to measure explicit frame-dependence for relative tensors, as in
the case here for the relative vorticity, is to directly compare
the two vorticity expressions in each frame separately:
\begin{equation}
\left.
\begin{aligned}
\text{For a relative tensor we face explicit frame-dependence, if
$\vW(\vL_{\vOmega}(\vx))\neq\vW(\vL(\vx))$},&\\[0.0em]
\text{or equivalently, if
$\tilde{\vW}(\tilde{\vL}_{\tilde{\vOmega}}(\tilde{\vx}))
\neq\tilde{\vW}(\tilde{\vL}(\tilde{\vx}))$.}&
\end{aligned}
~~~\right\} \label{161021:2226}
\end{equation}
The standard approach to measure frame-dependency as introduced in
Section~\ref{Sec2}, namely to verify whether
$\tilde{\vW}(\cdot)\neq\vW(\cdot)$ holds or not, only works for
absolute tensors, but not for relative tensors where this approach
degenerates to an action which only measures the relative
dependency between two different (non-inertial) frames
$\tilde{\vOmega}$ {\it and} $\vOmega$, and not the wanted
frame-dependency within one (non-inertial) frame $\tilde{\vOmega}$
{\it or} $\vOmega$. Hence, obviously, the relative vorticity only
shows a relative frame-indifference (relative objectivity), since
when evaluating the relative tensor relation~\eqref{161021:1544}
by using the definition of the vorticity $\vW$~\eqref{161013:1636}
and the transforation rule for the relative velocity gradient
$\vL_{\vOmega}$ \eqref{161019:0943}, one obtains the invariance
\begin{align}
\tilde{\vW}(\tilde{\vL}_{\tilde{\vOmega}}(\tilde{\vx}))
&=\vQ\vW(\vL_{\vOmega}(\vx))\vQ^T
=\vQ\left({\textstyle\frac{1}{2}}\big(\vL_{\vOmega}(\vx)
-\vL_{\vOmega}(\vx)^T\big)\right)\vQ^T\nonumber\\[0.5em]
&=\vQ\left({\textstyle\frac{1}{2}}
\big(\vQ^T\tilde{\vL}_{\tilde{\vOmega}}(\tilde{\vx})\vQ
-\vQ^T\tilde{\vL}_{\tilde{\vOmega}}(\tilde{\vx})^T\vQ\big)\right)\vQ^T
={\textstyle\frac{1}{2}}
\big(\tilde{\vL}_{\tilde{\vOmega}}(\tilde{\vx})
-\tilde{\vL}_{\tilde{\vOmega}}(\tilde{\vx})^T\big)
\nonumber\\[0.5em]
&=\vW(\tilde{\vL}_{\tilde{\vOmega}}(\tilde{\vx})),\;\;\text{i.e.}\;\;
\tilde{\vW}(\cdot)=\vW(\cdot),
\end{align}
which explicitly shows that the relative vorticity tensor is an
objective quantity only in the relative sense, since
$\tilde{\vW}(\cdot)=\vW(\cdot)$, but, as discussed before, not in
the absolute sense, due to its explicit frame dependence
$\vW(\vL_{\vOmega}(\vx))=\vW(\vL(\vx))+\vOmega\neq \vW(\vL(\vx))$.

\section{Classical vs. extended thermodynamics from the view of~MFI\label{SecD}}

In a nutshell, the difference in the modelling procedure of
classical and extended thermodynamics, described in detail in
\cite{Mueller98,Jou10,Siginer14,Ruggeri15}, boils down to the
following: Extended thermodynamics extends the number of fields
beyond the ones of ordinary or classical thermodynamics which are
mostly the densities of mass, momentum and energy. Typical
extensions are the fluxes of momentum and energy. A simple way to
obtain the evolution equations for these fluxes from a macroscopic
basis is to generalize the classical thermodynamic procedure,
however, always strictly in accord with the two incontrovertible
principles of entropy and relativity, namely to satisfy the
entropy balance law with a non-negative entropy
production,\footnote[2]{If the theory is not consistent with the
second law of thermodynamics, dissipative phenomena cannot be
modelled correctly.} and to ensure that all laws have the same
form and all processes induced by these laws run in the same way
in all inertial systems, either Galilean or Lorentzian, depending
on whether the theory is non-relativistic or relativistic.

Hence, in extended thermodynamics the fluxes are no longer
considered as mere control parameters but as own independent
variables, thus enlarging the range of applicability of
non-equilibrium thermodynamics to a vast domain of phenomena where
memory, non-local, and non-linear effects are relevant. In
particular for high-frequency phenomena the independent character
of the fluxes is made evident by the fact that the fluxes are fast
variables that decay to their local-equilibrium values after a
short relaxation time, thus allowing to describe phenomena at
(high) frequencies comparable to the inverse of the (low)
relaxation times of the fluxes. Also the aspect that the field
equations of ordinary thermodynamics are parabolic while extended
thermodynamics is governed by hyperbolic systems allowing only
finite speeds of propagation, shows that the procedure of extended
thermodynamics is superior over the ordinary or classical approach
of thermodynamics for irreversible processes.

Not summarized so far is the difference between these two
approaches concerning the principle of material frame-indifference
(MFI). For that, however, a closer look is necessary, which I
briefly want to illustrate in terms of a generic example without
specifying physical details, in order to bring out the essence of
the modelling procedure in each case from the perspective of~MFI:
In classical thermodynamics (for irreversible processes) the aim
is to determine the space-time evolution of some fields, say
$\vphi=(\phi_1,\ldots,\phi_n,\ldots \phi_N)_{1\leq n\leq N}$, also
called densities, satisfying a set of unclosed balance~equations
\begin{equation}
\frac{\partial \vphi}{\partial t}+\nabla\cdot
\vF=\vPi,\label{161025:2335}
\end{equation}
where the quantities $\vF$ are called the fluxes and $\vPi$ the
productions; in classical thermodynamics the latter are mostly
predetermined and thus known quantities of the system which can be
expressed in terms of $\vphi$ and $\vF$ and their gradients. If
the balance law \eqref{161025:2335} for some component $\phi_n$
represents a conservation law, then the corresponding production
$\Pi_n$ vanishes globally, if not, then $\Pi_n$ only vanishes
locally in all states of equilibrium. In order to solve for the
densities $\vphi$, the above unclosed equations of balance must be
supplemented by constitutive relations which relate the unknown
fluxes $\vF$ to the densities $\vphi$ in a manner characteristic
for the material considered. In classical thermodynamics, when
employing a microscopic description based on the Boltzmann
equation for some microscopic interaction model (mostly Maxwellian
molecules), the constitutive relations for the fluxes will
have~the~form
\begin{equation}
\vF=\vF^{(c)}(\vphi,\nabla\vphi,\nabla^2\vphi,\vGamma),\label{161025:2323}
\end{equation}
when applying the iterative Maxwell scheme$\,$\footnote[2]{See
footnote on p.$\,$\pageref{N2}.} up to second order (which mostly
suffices as a good approximation to the exact values) --- for
example, for the fluxes of momentum and energy the first order
Maxwell iterates represent the classical constitutive equations of
Navier-Stokes and Fourier. The result \eqref{161025:2323} thus
depends at each point on the values of the fields at that point
and on their values in the immediate neighbourhood, symbolically
denoted as the gradients $\nabla\vphi$ and $\nabla^2\vphi$, but
also on all parameters collectively denoted as $\vGamma$
characterizing the state of the material, either if passively
observed within a non-inertial frame of reference or if actively
transformed into a new non-inertial state.\footnote[3]{The
explicit frame-dependency $\vGamma$ in the iterative result
\eqref{161025:2323} is essentially picked up through the balance
equations \eqref{161025:2335} themselves when formulated for, or
transformed to any non-inertial system. For inertial systems or
transformations connecting inertial systems, however, all
frame-dependency parameters $\vGamma$ will vanish if the set of
all balance equations \eqref{161025:2335} show frame-indifference
under the inertial Galilei transformations (or Lorentz
transformations in relativistic continuum mechanics), which for
closed systems is a strict principle of nature and must always be
satisfied --- where here it is to note that the requirement
`closed system' need not to be closed in the thermodynamic sense,
which obviously would be too restrictive. For example, the
classical Navier-Stokes equations without external body forces are
frame-indifferent and thus fully symmetric under Galilei
transformations, although they are not thermodynamically closed
due to their dissipative nature. Hence, under a `closed system'
regarding frame-indifference, we will henceforth only understand
the much weaker requirement of no external body forces acting on
the thermodynamic system, as such forces may break the space-time
symmetries necessary for Galilean or Lorentzian invariance; for
example, a spatially and temporally constant external force breaks
the isotropy of~space.\label{N3}} Hence, it is in this sense that
MFI is violated as expressed in \cite{Mueller72,Mueller76}. In
other words, to demand frame-indifference for the constitutive
relations~\eqref{161025:2323} within the procedure of ordinary
thermodynamics would obviously be wrong.

In extended thermodynamics, however, the number of fields is
extended by identifying the constitutive flux $\vF$ in
\eqref{161025:2335} as a density satisfying its own balance
equation, thus augmenting the set of equations \eqref{161025:2335}
of classical thermodynamics with a new set of equations of higher
order
\begin{equation}
\left.
\begin{aligned}
\frac{\partial \vphi}{\partial t}+\nabla\cdot \vF=\vPi &,\\[0.5em]
\frac{\partial \vF}{\partial t}+\nabla\cdot \vJ=\vXi
&,\label{161025:2345}
\end{aligned}
~~~~~ \right\}
\end{equation}
where $\vJ$ and $\vXi$ are, respectively, the new (unclosed)
fluxes and productions of the combined system $(\vphi,\vF)$, i.e.,
in contrast to the productions $\vPi$ of classical thermodynamics,
the productions $\vXi$ of extended thermodynamics are unknowns of
the system which need to be modelled, too. Thus in order to solve
for these densities, the balance equations \eqref{161025:2345}
must be supplemented again by constitutive relations, which now,
in the extended version, have to relate $\vJ$ and $\vXi$ to the
densities $(\vphi,\vF)$. In the framework of extended
thermodynamics, all constitutive relations of the augmented system
\eqref{161025:2345} are modelled locally and instantaneously in
space-time~as
\begin{equation}
\vJ=\vJ^{(e)}(\vphi,\vF),\qquad
\vXi=\vXi^{(e)}(\vphi,\vF),\label{161026:1038}
\end{equation}
so that the fluxes $\vJ$ and the productions $\vXi$ at a point and
a time depend only on the densities $(\vphi,\vF)$ at that point
and time, in contrast to the form of the constitutive relations
\eqref{161025:2323} in ordinary thermodynamics, which standardly
depend also on the gradient of the densities. As explained and
discussed in detail in \cite{Mueller98} and \cite{Ruggeri15}, the
local and instantaneous modelling ansatz \eqref{161026:1038} is
superior to any other ansatz as this naturally leads to a
thermodynamic system which is governed by hyperbolic field
equations allowing only finite speeds of propagation, in contrast
to the field equations of ordinary thermodynamics which, mainly
due to the non-instantaneous structure of the fluxes
\eqref{161025:2323}, are in general of parabolic type allowing for
unphysical propagations of infinite speed.

Now, with the aim to see how extended thermodynamics relates to
classical thermodynamics, it is necessary to construct from the
augmented balance equation \eqref{161025:2345} a constitutive
relation for the fluxes $\vF$ of the classical type
\eqref{161025:2323}. This relation is provided again by a formal
iterative scheme akin to the Maxwell scheme of classical
thermodynamics: The first iterates of the fluxes $\vF$ are
obtained from the right-hand sides of the extended balance
laws~\eqref{161025:2345} by putting the equilibrium values into
the left-hand sides, while the second iterates are again obtained
from the right-hand sides but now by putting the first iterates
into the left-hand sides, and so on. Ultimately this iteration can
be considered as a power expansion of the thermodynamic relaxation
times of the fluxes $\vF$, where with each iterate the
thermodynamic state is described further and further away from
equilibrium.\footnote[2]{Note that in a far-from-equilibrium case
it is still a major open problem whether the solutions via
Maxwellian iterations converge or not. A rigorous proof for
convergence is still missing \citep{Ruggeri15}.} For small
relaxation times, however, i.e. close or not far from equilibrium,
the iteration can already be terminated at second order to yield a
sufficiently good approximation for the fluxes, giving then the
comparative result to \eqref{161025:2323} of classical
thermodynamics
\begin{equation}
\vF=\vF^{(e)}(\vphi,\nabla\vphi,\nabla^2\vphi,\vGamma)
\,\sim\,\vF^{(c)}(\vphi,\nabla\vphi,\nabla^2\vphi,\vGamma),\label{161027:1738}
\end{equation}
showing that classical thermodynamics is only a valid theory close
to equilibrium for small relaxation times. In other words,
extended thermodynamics not only specifically shows that classical
constitutive relations as \eqref{161025:2323}, which (due to the
appearance of gradients) are non-local in space, are
approximations of some higher-order balance laws involving only
certain spatially local constitutive relations
\eqref{161026:1038}, but also more generally that parabolic
systems of classical theories in principle appear as
approximations of corresponding higher-order hyperbolic systems
when some relaxation times are negligible.

When regarding the problem of MFI, however, the above procedure of
extended thermodynamics shows the striking feature that while the
Maxwell iterate \eqref{161027:1738} for the fluxes $\vF$ picks up
again the frame-dependency $\vGamma$ when the balance laws
\eqref{161025:2345} are formulated or transformed to a
non-inertial system, the constitutive equations of the extended
fluxes and productions \eqref{161026:1038}, however, remain
frame-indifferent. Not because of the fact that they do not
transform, but because of the fact that they are modelled as such.
Any frame-dependency $\vGamma$ is manually eliminated from
\eqref{161026:1038} by only allowing for frame-indifferent
operators and fields, thus constituting a strong restriction on
the possible functional structures of \eqref{161026:1038}. For
example, for non-inertial (time-dependent) rotations, the
structure of \eqref{161026:1038} is restricted to isotropic
functions.

However, as explained and discussed in Section \ref{Sec4.2}, to
restrict the extended constitutive relations \eqref{161026:1038}
according to MFI and to proclaim it as an axiom of nature, is not
convincing. For an appropriate axiom in physics to model
form-invariantly (covariantly) and frame-indifferently
(objectively) in accordance with all physical observations
existing, the general principle of MFI, which is based on the
non-inertial Euclidean transformations, has to be replaced by the
reduced principle r-MFI (as formulated in Section \ref{Sec4.3})
restricting constitutive modelling only according to the inertial
Galilean transformations. For particular modelling cases, however,
as turbulence modelling, where the microscopic description is
exactly known, the axiom r-MFI can be naturally supplemented by
including additional modelling restrictions as worked out in
Section \ref{Sec5}.

\bibliographystyle{jfm}
\bibliography{BibDaten}

\end{document}